\documentclass[12pt,preprint]{aastex}

\slugcomment{March 24 -- to appear in ApJS}

\shorttitle{Rotational modulation of X-rays in the ONC}
\shortauthors{Flaccomio et al.}

\begin{document}

\title{Rotational modulation of X-ray emission in Orion
Nebula young stars}

\author{E. Flaccomio\altaffilmark{1}, G. Micela\altaffilmark{1},                   S. Sciortino\altaffilmark{1}, E. D. Feigelson\altaffilmark{2},
          W. Herbst\altaffilmark{3}, F. Favata\altaffilmark{4},
          F.R. Harnden Jr.\altaffilmark{5,6} and  S. D.     
	  Vrtilek\altaffilmark{5}}

\altaffiltext{1}{INAF-Osservatorio Astronomico di Palermo Giuseppe S. Vaiana,
Piazza del Parlamento 1, 90134 Palermo, Italy}
\email{ettoref@astropa.unipa.it}

\altaffiltext{2}{Department of Astronomy and Astrophysics, 525 Davey
Laboratory, Pennsylvania State University, University Park, PA 16802, U.S.A}

\altaffiltext{3}{Department of Astronomy, Wesleyan University, Middletown, CT
06459, U.S.A}

\altaffiltext{4}{Astrophysics Division, Research and Science Support Dept. of
ESA, Postbus 299, 2200 AG Noordwijk, The Netherlands}

\altaffiltext{5}{Smithsonian Astrophysical Observatory, 60 Garden St.,
Cambridge, MA 02138, U.S.A}

\altaffiltext{6}{Universe Division, Science Mission Directorate NASA
Headquarters}

\begin{abstract}
\end{abstract}

We investigate the spatial distribution of X-ray emitting plasma in a
sample of young Orion Nebula Cluster stars by modulation of their X-ray
light-curves due to stellar rotation. The study, part of the {\em
Chandra Orion Ultradeep Project} (COUP), is made possible by the
exceptional length of the observation: 10 days of ACIS integration
during a time span of 13 days, yielding a total of 1616 detected
sources in the $17\times17$ arcmin field of view. We here focus on a
subsample of 233 X-ray-bright stars with known rotational periods. We
search for X-ray modulation using the Lomb Normalized Periodogram
method.

X-ray modulation related to the rotation period is detected in at least
23 stars with periods between 2 and 12 days and relative amplitudes
ranging from 20\% to 70\%. In 16 cases, the X-ray modulation period is
similar to the stellar rotation period while in seven cases it is about
half that value, possibly due to the presence of X-ray emitting
structures at opposite stellar longitudes. These results constitute the
largest sample of low mass stars in which X-ray rotational modulation
has been observed. The detection of rotational modulation indicates
that the X-ray emitting regions are distributed inhomogeneneously in
longitude and do not extend to distances significantly larger than the
stellar radius. Modulation is observed in stars with saturated activity
levels ($L_X/L_{bol} \sim 10^{-3}$) showing that saturation is not due
to the filling of the stellar surface with X-ray emitting regions.


\section{Introduction \label{sect:intro}}

Pre-main Sequence (PMS) stars have high levels of X-ray activity with
non-flaring X-ray luminosities up to $10^{31}\ \rm erg\ s^{-1}$
\citep{pre05}. As in the case of older cluster and field stars, X-ray
activity in the PMS phase has often been attributed to a ``scaled up''
solar-like corona formed by active regions but more densely packed on
the stellar surface and/or having higher plasma densities than on the
Sun. For most non-accreting PMS stars (Weak-line T-Tauri Stars, WTTS),
the fraction of energy emitted in the X-ray band with respect to the
total stellar output, $L_{\rm X}/L_{\rm bol}$, is close to the
saturation level, $10^{-3}$, seen on rapidly rotating main sequence
(MS) stars \citep{fla03b,pre05,piz03}.  This suggests a common 
physical mechanism for the emission of X-rays or, at least, for its
saturation in PMS and MS stars.

However, the solar analogy, while providing a simple picture of
activity in PMS stars, may not be fully valid.  Saturated PMS stars
have $L_{\rm X}/L_{\rm bol}\sim1000$ times greater than the Sun at its
maximum, when the solar surface is $\sim$50\% covered with active
regions and $\sim$1\% covered with active regions cores
\citep{dra00,orl01}. Plasma temperatures in saturated stars are much
higher than on the Sun and, in the few cases where surface
distributions can be determined, their active regions are concentrated
at higher latitudes (e.g. \citealt{fav99}, \citealt{jard02}). The
nature of the saturation phenomenon is not understood. It might be due
to filling of the stellar surface with active regions, saturation of
the magnetic reconnection process that heats the plasma, centrifugal
limits on the coronal extent, concentration of magnetic fields towards
the poles, or saturation of the internal dynamo generating the magnetic
field that confines the plasma \citep{Gudel04}.

Even more puzzling than WTTS are PMS stars that are still undergoing
mass accretion (Classical T-Tauri Stars, CTTS). CTTS, with their
circumstellar disks and magnetically funneled matter inflows and
outflows host in general several different physical phenomena. With
respect to X-ray activity, the bulk of the observational evidence
points toward phenomena similar to those  occurring on WTTS. However,
CTTS have significantly lower and unsaturated values of $L_{\rm X}$ and
$L_{\rm X}/L_{\rm bol}$ \citep{dami95,fla03a,fla03b,pre05}.  Similar to
saturated WTTS, their X-ray activity does not correlate with stellar
rotation, as seen in unsaturated MS stars and naturally interpreted in
terms of a rotationally regulated $\alpha-\Omega$ dynamo. A further
complication is that the high-resolution X-ray spectra of two observed
CTTS, TW Hya \citep{kast02,stel04} and BP Tau \citep{sch05}, indicate
that a fraction of the soft X-ray luminosity may be produced in
accretion shocks rather than coronal magnetic reconnection events.

In contrast to emission close to the stellar surface, some of the
strongest flares seen in PMS stars are most easily explained by
magnetic loops connecting the stellar surface with a circumstellar disk
\citep{fava05}. Such loops arise naturally in models of magnetically
driven accretion that seem to best explain the shapes of optical
emission lines \citep{shu94,muze01}.  Magnetic reconnection in these
loops might be the source of powerful and long-lasting X-ray flares
\citep[e.g.][]{shu97,mont00}.  The influence of circumstellar disks
and/or accretion on the emission of X-rays is not altogether unexpected
given that the stellar magnetic field responsible for the confinement
of X-ray emitting plasma is likely affected by the circumstellar
environment.

Observations of rotationally induced modulation of X-ray emission from
low mass stars can potentially yield valuable information on the extent
and spatial distribution of the magnetic structures in active stars.
Such observations are to date very scarce. X-ray rotational modulations
have been seen in VXR 45, a fast rotating ($P_{rot}=0.22$d) G9 ZAMS
star \citep{mar03}, AB Dor \citep{huss05}, a K0 ZAMS star with
$P_{rot}=0.5$d, and possibly EK Dra ($P_{rot}=2.7$d) \citep{gue95}. To
the authors' knowledge, X-ray rotational modulation has not been to
date reported for any PMS star. Because of the typical duration of
X-ray observations, $\sim 1$ day, it is indeed difficult to detect
rotational modulation in the majority of stars which have much longer
periods.

The $Chandra$ Orion Ultradeep Project (COUP) has detected over 1616
sources in an 850 ks exposure spanning 13.2 d of the rich young Orion
Nebula Cluster \citep[ONC][]{get05}, well in the range of optically
determined periods of PMS stars \citep{herb02}. With COUP, we thus have
the unique opportunity to study rotational modulation of X-ray emission
on a large sample of PMS stars. We focus here on COUP sources whose
rotation period is known from optical studies. Results from the
analysis method discussed below are also presented for specific COUP
sources by \cite{stel05} and \cite{gros05}.

The paper starts with a definition of the catalog of COUP sources with
known rotation periods (\S \ref{sect:Popt}). Section \ref{sect:dataana}
describes the X-ray light curves and the analysis method used to
estimate X-ray modulation periods and their reliability. Results of the
period search are presented in \S \ref{sect:results}, followed by
discussion of implications for the physical origin of X-ray activity in
PMS stars (\S \ref{sect:discussion}). Finally in \S \ref{sect:summconc}
we summarize our results.

\section{Catalogs of optical periods}
\label{sect:Popt}

Our primary goal is to establish whether rotational modulation of X-ray
emission is observed in COUP sources. Our initial sample comprises 295
COUP sources for which we could find in the literature a rotational
period ($P_{\rm opt}$) determined from the periodic modulation of
optical or near-infrared light curves.

We started with the list of 201 optical periods measured by
\cite{herb02} and reported by \cite{get05} for COUP sources.  Two of
these COUP sources, 150 and 1328, are identified with close pairs: COUP
150 with sources  010-411 and 011-410 in \cite{luc00} and with sources
201 and 10342 in \cite{herb02}\footnote{Entries for source 201 and
10342 in \cite{herb02} are very similar and they might actually refer
to the same object, but only 201 has a rotation period. Source 010-411
and 011-410 in \cite{luc00} are 1.1\arcsec\/ apart (seeing
$\sim0.6$\arcsec) and have similar magnitudes and colors. The H
magnitudes add up to the 2MASS value of 10.6, but the J magnitudes add
up to 10.05 while 2MASS gives 11.48.}, of which only 201 has a rotation
period. COUP 1328 is associated with both 776a (with measured rotation
period) and 776b in \citet[hereafter H97]{H97}. In both cases,
following \cite{get05}, we associated the X-ray source with the
rotation period measured by \cite{herb02} for one of the components.
Our search for modulation in the X-ray light curves yields periods that
appear related to these optically determined periods, thus confirming
the association. We exclude the rotation period listed by \cite{get05}
for COUP 460, associated with source 347 in the catalog of
\citet[hereafter JW]{jones88}, for which we could not find an optical
period in the literature. Furthermore, we have updated the optical
periods for COUP 1432 (JW 843, P=5.38d), 1443 (JW 852, P=1.32d) and
1487 (JW 883, P=5.70d), based on unpublished improved determinations by
W. Herbst.

We then added periods from recent literature: 51 from \cite{herb00}; 15
from \cite{stas99}; 23 from the list of \cite{carp01} (selecting only
periods with FAP $<$ 1\%); 3 from \cite{herb02} missed by \cite{get05}
(COUP 3 = JW 25, P=2.28d; COUP 68 = JW 130, P=1.2d; COUP 252 = JW 275a,
P = 0.95d); and 3 unpublished by W.\ Herbst (COUP 328 = JW 330,
P=4.08d; COUP 1350 = JW 795, P=2.80d; COUP 1154 = JW 684, P=3.35d).

Table \ref{tab:sample} lists the 295 COUP sources that are thus
assigned a rotational period. In cases when more than one optical
period was retrieved from the literature, we list other values if
differing by more than 10\% from the adopted one.

\section{X-ray light curves and analysis method}
\label{sect:dataana}

Adopting the source photon extraction of the ACIS X-ray data presented
in \cite{get05}, we bin arrival times and compute the count rate in
each bin taking into account gaps in the observation. We then subtract
the mean background count rate\footnote{For all sources the background
rate is statistically indistinguishable from being constant within the
observation.} as determined from a source-free region close to each
source \citep{get05}. The period analysis described below  was
originally performed wi1th bin lengths of 10000, 5000, and 2000
seconds. Results obtained in the three cases were very similar, both
regarding the modulation periods found and their statistical
significances, though slightly better significances were obtained with
shorter bin lengths, probably as a result of an improved ability to
account for non-rotationally induced variability (e.g. flares). In the
following, we therefore refer exclusively to the analysis performed
using 2000 seconds bins. 


Our goal is to detect rotational modulation in the X-ray light curves
of COUP sources. We have initially considered two period analysis
methods: the Lomb Normalized Periodogram (LNP) method \citep{scar82}
and the string length method (SLM) with length defined as in
\citet{lafl65}, but soon realized that the former gave better results.
The LNP is the traditional \citet{Shuster1898} periodogram of Fourier
analysis adjusted for gaps in the data: a function is computed from the
data for a range of test frequencies and the frequency that maximizes
this function is considered as the most likely frequency in the data.

\begin{figure}[!t]
\centering
\includegraphics[width=8cm,clip=true]{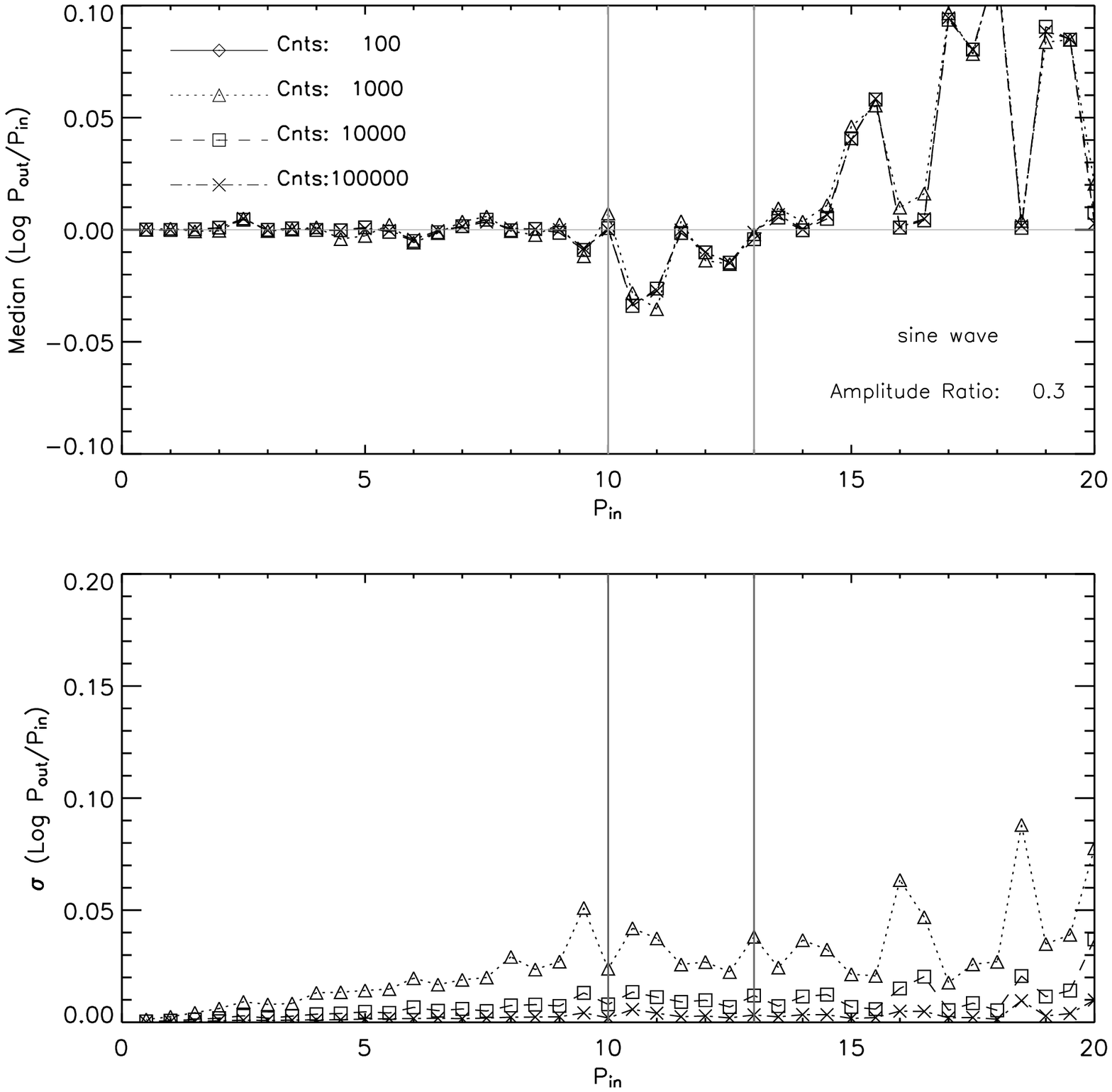}
\includegraphics[width=8cm,clip=true]{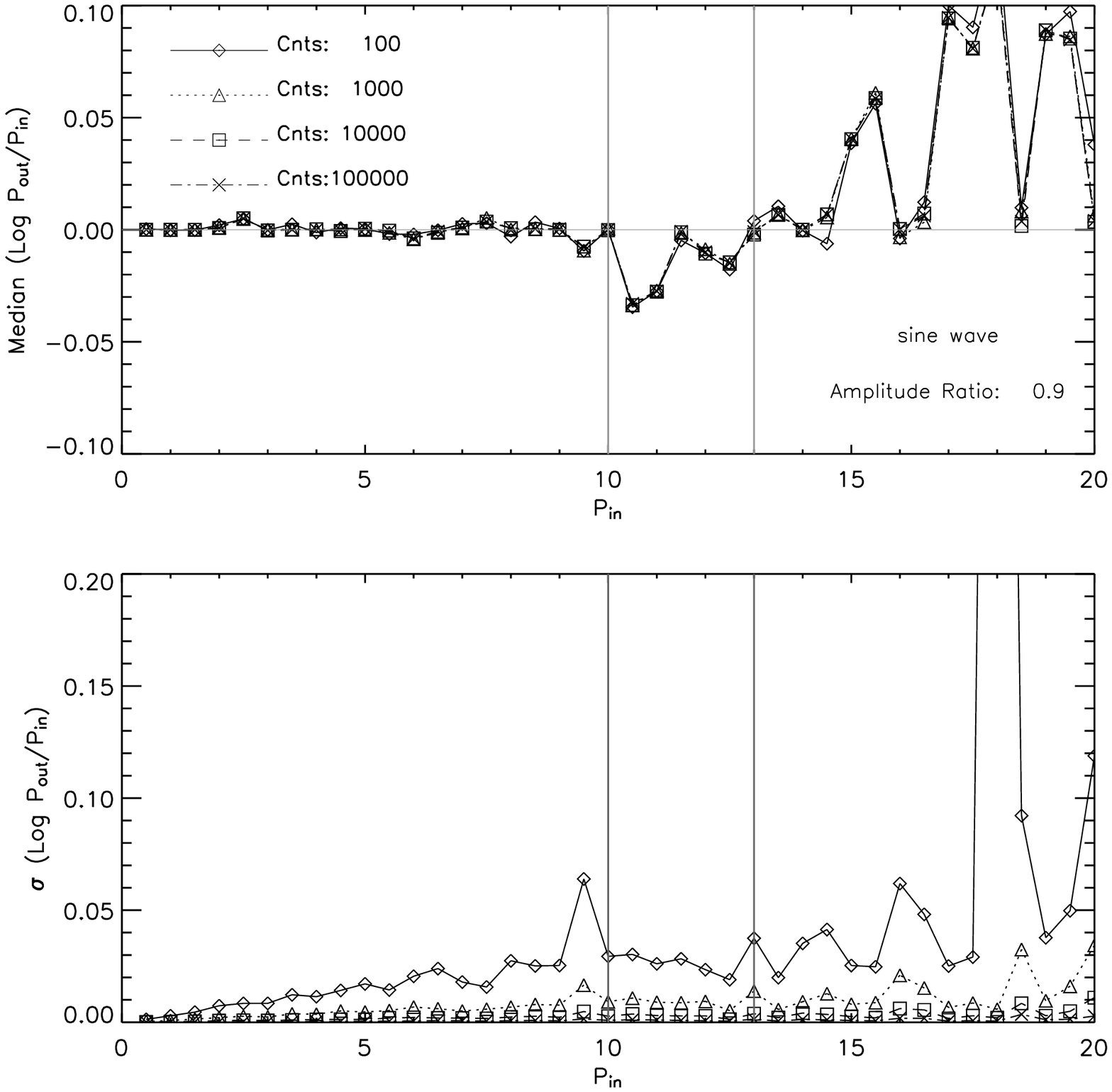}

 \caption{Results of simulations of purely periodic sinusoidal light
curves for relative modulation amplitudes of 30\% (left) and 90\%
(right). Upper panel: log median $P_{in}/P_{out}$, as a function of
$P_{in}$; $P_{in}$ is the input period for the simulation and $P_{out}$
is the period corresponding to the main peak in the resulting
periodogram. Each point is the result of 100 simulated sinusoidal light
curves with counts ranging from 100 to 100000, as indicated in the
label. Vertical lines indicate periods of 10 and 13 days, i.e. the
exposure time and the total length of our observation including gaps.
Lower panels: 1 $\sigma$ scatter of the 100 simulation around the
median for the same simulations. \label{fig:simul}}

\end{figure}

In order to estimate the range of applicability of the LNP analysis to
our data we first performed a series of simple simulations by
generating purely periodic light curves in $2000\ s$ bins with the
observed window function. Three lightcurve shapes were considered: a
sinusoid ($M[1+A\sin \omega t]$), a three level ($M[1-A],M,M[1+A]$)
step function, and a two level ($M[1-A],M[1+A]$) step function.
Simulated periods ranged from 0.5 to 30 days and relative amplitudes
($A$) from 10\% to 90\%. Noise was assumed Poissonian, appropriate for
total source counts ranging from 100 to $10^5$. One hundred replicates
were run for each set of input parameters. Figure \ref{fig:simul} shows
the results for the cases of sinusoidal light curves and $A=30$\% and
$A=90$\%. The plots show the median and the 1$\sigma$ dispersion of the
ratio between the period corresponding to the main LNP peak ($P_{out}$)
and the input period $P_{in}$ as a function of input period and
simulated source counts. Note that, at this stage, we are not yet
estimating significances for the  peaks in the periodograms as in the
more sophisticated simulations discussed in Sect. \ref{sect:fsim}.  The
simulations here indicate that a certain fraction of the $P_{out}$ will
be unrelated to $P_{in}$ when the statistics and/or the amplitudes are
too low. In the $A=30$\% case, for example, the curves referring to the
100 count lightcurves lie outside the plot vertical scale. For high
count rates, the median $P_{out}/P_{in}$ does not diverge by more than
0.03 dex for $P_{in}<13$ days, the length of the observation (including
gaps in the data stream). For simulations with 100 (1000) counts, the
minimum amplitude that still yields $P_{in}/P_{out} \sim 1$ for $P_{in}
< 13$ d is $A\sim60$\% (20\%). Similar results are obtained if the
shape of the simulated modulation is changed from a sinusoid to a 2 or
3 level step function.

Given these results, in the analysis of real data we therefore decided
to limit our search to X-ray periods $P_X$ shorter than 13 days. The
lower limit to the accessible $P_X$ is set by timescales of the
non-rotational variability present in our data, notably flares. Because
of variability unrelated to rotation, the simulation discussed above
represent a best case scenario and can only indicate a lower limit to
the typical uncertainties in the determination of periods. In Sect.
\ref{sect:fsim}, after presenting the results of the LNP analysis of
COUP sources, we introduce more realistic simulations of light curves
that include flares.

\subsection{Calculation of FAPs and filtering of light curves}
\label{sec:analysis}

The most difficult part of the periodogram analysis in the non-ideal
case is establishing the significance of the peaks found in the LNP or,
more specifically, the false alarm probabilities (FAPs). Stochastic
noise and intrinsic variability, unrelated to rotational modulation can
produce spurious features in the LNP that can easily be mistaken for
real modulation. A simple example, which recurs frequently in our data,
is that of repeated flares. Two bright flares occurring on the same
source within our observation with a temporal separation $\Delta T$
will produce a spurious peak in the periodogram at frequency $1/\Delta
T$. It is therefore paramount to either take into account the
non-rotationally induced variability in the calculation of the FAPs
and/or to filter out identifiable features from the light curves prior
to the calculation of the LNP.

In the presence of uncorrelated noise, FAPs can be computed from the
actual light curve through a permutation resampling technique. A large
number of artificial light curves are computed by replacing actual
count-rates with values randomly chosen from the same light curve. The
LNP analysis is then performed on each of these realizations and, for
each frequency, the distribution of power values is recorded. Once the
maximum from the real periodogram is found, its value can be compared
to the distribution obtained, at the same frequency, from the
randomized light curves in order to establish the probability that
values as high as the observed one are due to random fluctuations; this
is the FAP.

Performing this calculation for our X-ray light curves always results
in vanishingly small FAPs, indicating that LNPs of COUP sources present
peaks that cannot be explained by purely stochastic noise. This arises
because flares with a variety of characteristics are ubiquitous
\citep{wolk05,fava05}, as well as less well-understood temporal
phenomena. We therefore seek to compute FAPs by randomizing any
periodic component of the light curves but preserving these other
correlated features. We define blocks of adjacent temporal bins with
total length $\tau_{corr}$, and randomize the position of these blocks.
Our $\tau_{corr}$ should be longer than flare timescales and shorter
than rotational periods, but these two conditions cannot always be
simultaneously achieved. We choose $\tau_{corr}=8$ hours and
consequently limit our search of peaks in the periodograms to periods
longer than 1 day. We caution that the choice of $\tau_{corr}=8$ hours
is somewhat arbitrary and a number of longer flares are observed
\citep{fava05}. For this reason the FAPs that we derive are not
formally accurate, but we are confident they are good indicators of the
relative significance of the peaks in the periodograms.

An essential element of our analysis is to perform periodogram analysis
on both the original lightcurves and those where large flares are
removed. We filter the lightcurves using two trimming levels
established to qualitatively remove obvious flares: 1) a light trimming
(``L'') of bins with count rates higher than 1.5 times the 13\% upper
quantile of the count rate distribution; and 2) a heavy trimming
(``H'') of bins with count rates higher than 1.5 times the 25\%
quantile.  We also tried filtering out time interval that are
identified as belonging to flares using Maximum Likelihood Blocks as
described in \cite{wolk05}, but, except for a few sources, results were
not in general better than with the simpler count rate trimming
procedures.  The original lightcurves will be referred to as ``N'' (no
filter). Given the large variety of behaviors observed in our light
curves, we have not been able to identify a single best filtering
method for all cases. Moreover, our filtering may not always improve
our ability to find real modulation periods and to reject spurious
results due, for instance, to the alignment of untrimmed flare decay
tails.

Given the above considerations on the possible pitfalls involved in the
search for true modulation periods in the complex X-ray lightcurves of
COUP sources, we adopt the following graduated levels of confidence in
a possible rotational period detection: periodogram peaks are
considered most significant when FAP$<0.1$\%\footnote{FAPs were
computed with the bootstrapping procedure described in this section
using 100,000 iterations, sufficient to define the 0.1\% level} in all
three ("N", "L", and "H") lightcurves. Less confidence is given to
sources where FAP$<0.1$\% in any of the three lightcurves.  When more
than one filtering strategy exhibit a signal, we adopt the period with
the lowest FAP\footnote{In 7 cases more than one LNP peak has
FAP=0.0\%  within the precision of the determination. In these cases we
choose the highest LNP peak above the FAP=0.1\% threshold.}.

\section{Results} \label{sect:results}

Table \ref{tab:sample} lists optical and X-ray data for the 295 COUP
sources with known optical rotation periods.  The period values are
described in \S \ref{sect:Popt} and the remaining columns are obtained
from tables in \citet{get05}.  We performed the LNP period analysis for
233 of these sources; twenty-one sources with $<$100 X-ray photons and
41 sources with optical rotation period $P<2$ d were omitted from
analysis. This is to exclude stars for which period detection would
likely be hampered by low photon statistics and by the similarity
between the flare and modulation timescales. For the 233 analyzed
sources, the last column indicates whether a FAP$<0.1$\% peak is seen
in the periodogram of the `N', `L' and `H' lightcurves. One hundred and
nine sources are detected as periodic in at least one dataset (we call
this the ``$N+L+H$'' sample), 48 sources of these are detected as
periodic using unfiltered (the $N$ sample) lightcurves, and 34 of these
are the highest confidence periodic signals showing a significant peak
in all three datasets (the ``$N\cdot L\cdot H$'' sample). Detailed
results of the periodogram analysis for the $N\cdot L\cdot H$ sample
are given in Table \ref{tab:NLH}. For each source, we provide
modulation periods and FAPs for each of the three X-ray light curves.
The lowest FAP value (highest significance period) was obtained from
the unfiltered light curves in 4 sources, lightly filtered lightcurves
in 15 sources, and heavily filtered lightcurves in 15 sources (column
12 of Table \ref{tab:NLH}).

The 13th column in Table \ref{tab:NLH} reports an estimate for the
relative amplitude of the modulation from the folded X-ray lightcurve
derived from the lightcurve giving the lowest FAP. In order to reduce
random fluctuations when estimating amplitudes, we rebin the phased
lightcurve to give 8 bins per modulation cycle if the lightcurve has
less than 8000 counts and an average of 1000 counts per bin for
brighter sources. The relative amplitude is then defined as the
difference between the minimum and maximum count rates divided by the
sum of the same quantities. Resulting values are uniformly
distributed between $\sim 20$\% and $\sim$70\%.

\begin{figure*}[!t]
\includegraphics[width=17.0cm]{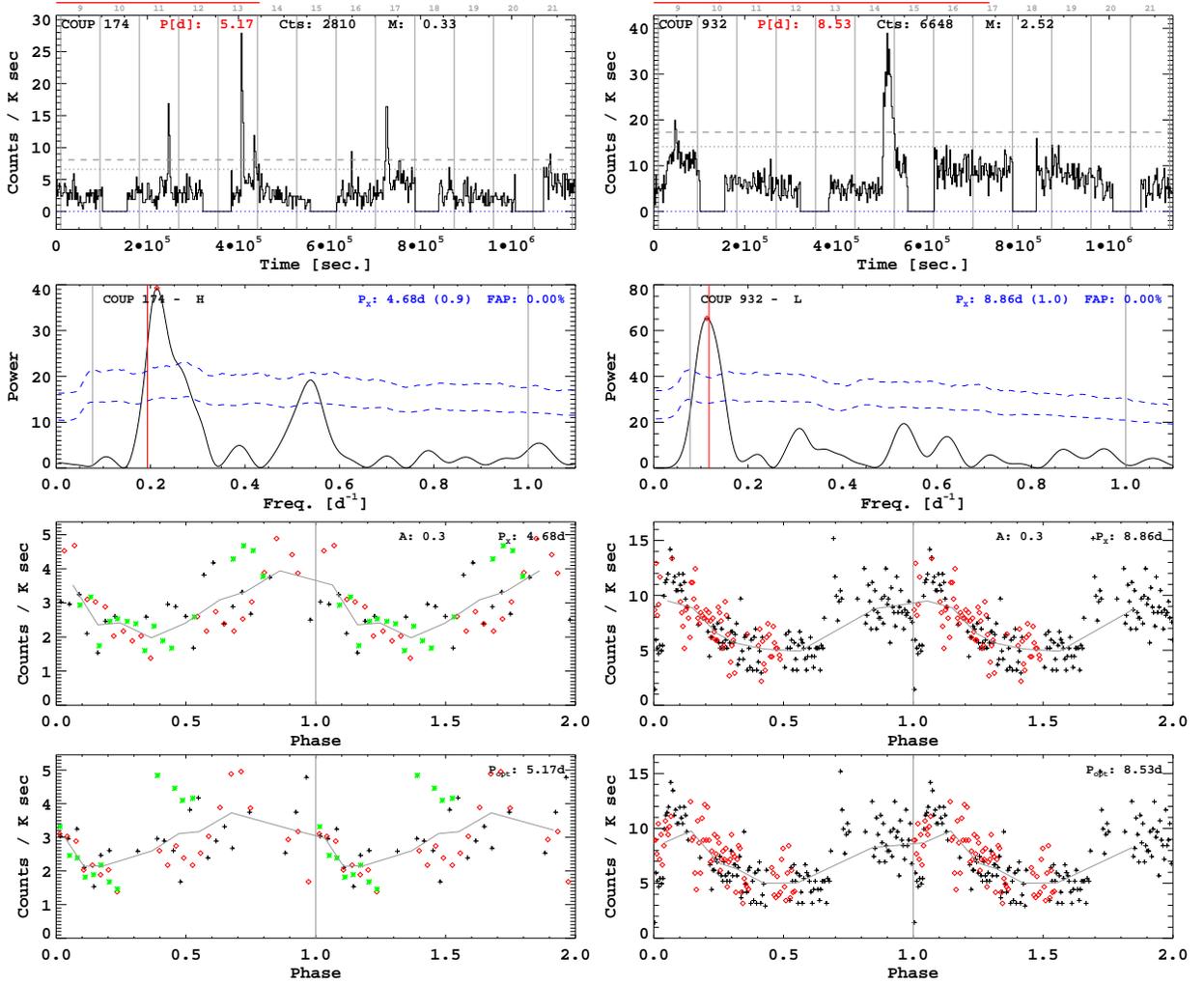}

\caption{Upper panels: light curves for two sources (COUP 174 and COUP
932) for which the modulation analysis yielded X-ray periods similar to
optically determined stellar rotation periods. Background subtracted
count rates in 2000s bins are plotted versus the time since the
beginning of the observation. UT dates (in January 2003) are given
above the plot, between gray vertical lines indicating UT midnights.
Source number, optical periods, extracted net counts and stellar mass
are given in the upper part of the plot. The red horizontal segment
above lightcurves indicates the length of one rotation period. The
dashed and dotted horizontal lines indicate count-rate thresholds used
for ``light trimming'' and ``heavy trimming'' respectively. Second row
of panels: Lomb normalized periodograms for the light curves filtered
so as to minimize the FAP. The filtering method (``N'' none, ``L'' light
trimming, ``H'' heavy trimming) is given next to the source id in the
upper right corner. The red vertical line indicates the frequency
corresponding to the optical rotation period. The two dashed curves
indicate 1\% and 0.1\% FAP thresholds (see text). The most likely X-ray
period is reported in the upper right corner (with in parenthesis the
ratio $P_X/P_{opt}$) along with the FAP. Third row: light curves folded
with the most likely X-ray period, reported in the upper right corner
along with the relative modulation amplitude. Different symbols and
colors indicate data-points belonging to different modulation cycles.
The solid gray line indicate the average count-rate as a function of
phase. Fourth row: light curves folded with the optically determined
rotation period. \label{fig:exLC0}}

\end{figure*}

\begin{figure*}[!t]

\includegraphics[width=17cm]{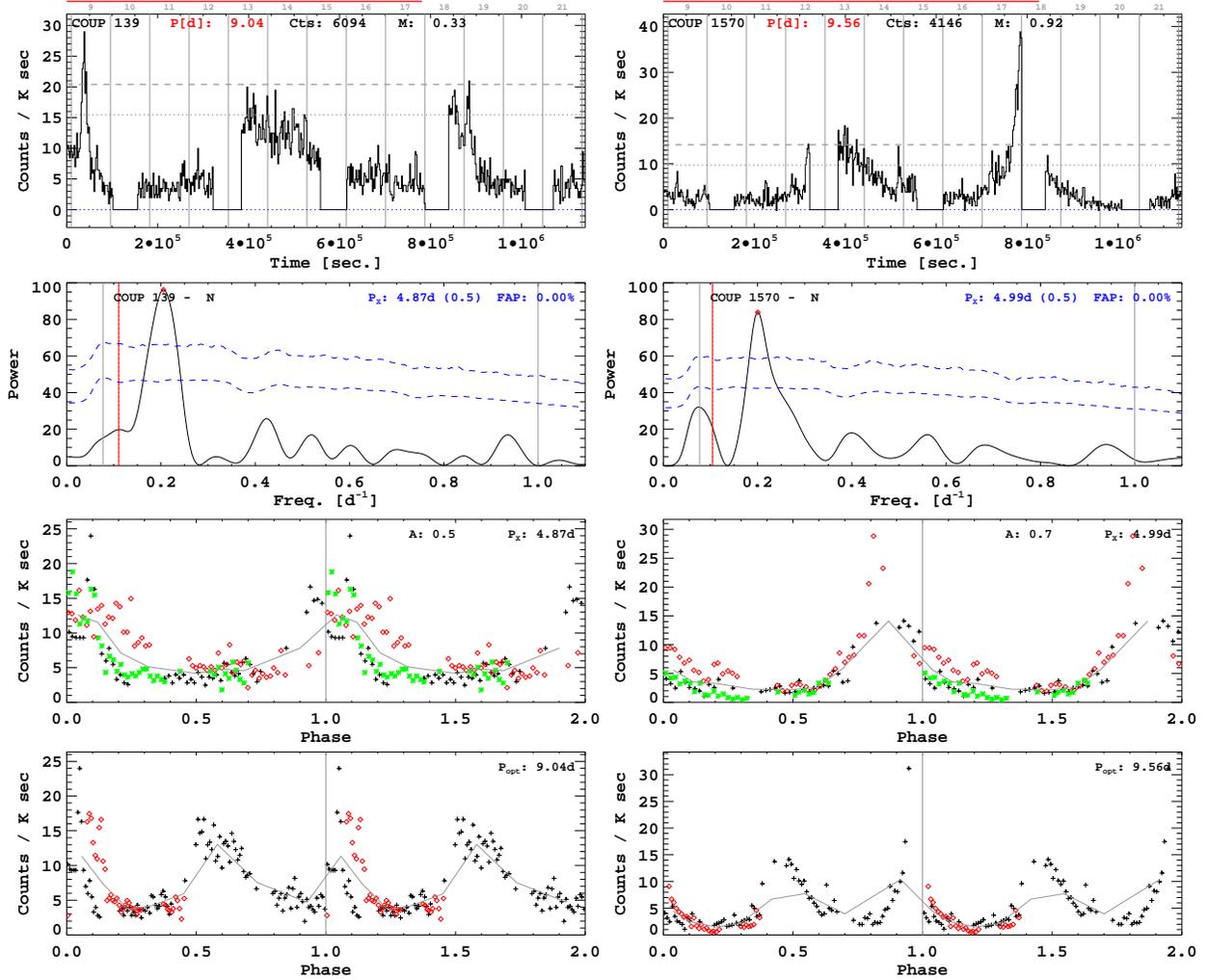}

\caption{Same as figure \ref{fig:exLC0} for two sources for which the
most likely X-ray period is half the optical period. Note however that
for the source on the left (COUP 139) a not significant peak is present
at the optical period. Applying the ``H'' filter, the peak at $P\sim 4.9$d
remains predominant but the FAP of this secondary peak drops to 0.7\%. 
\label{fig:exLC1}}

\end{figure*}

\begin{figure*}[!t]
\includegraphics[width=17cm]{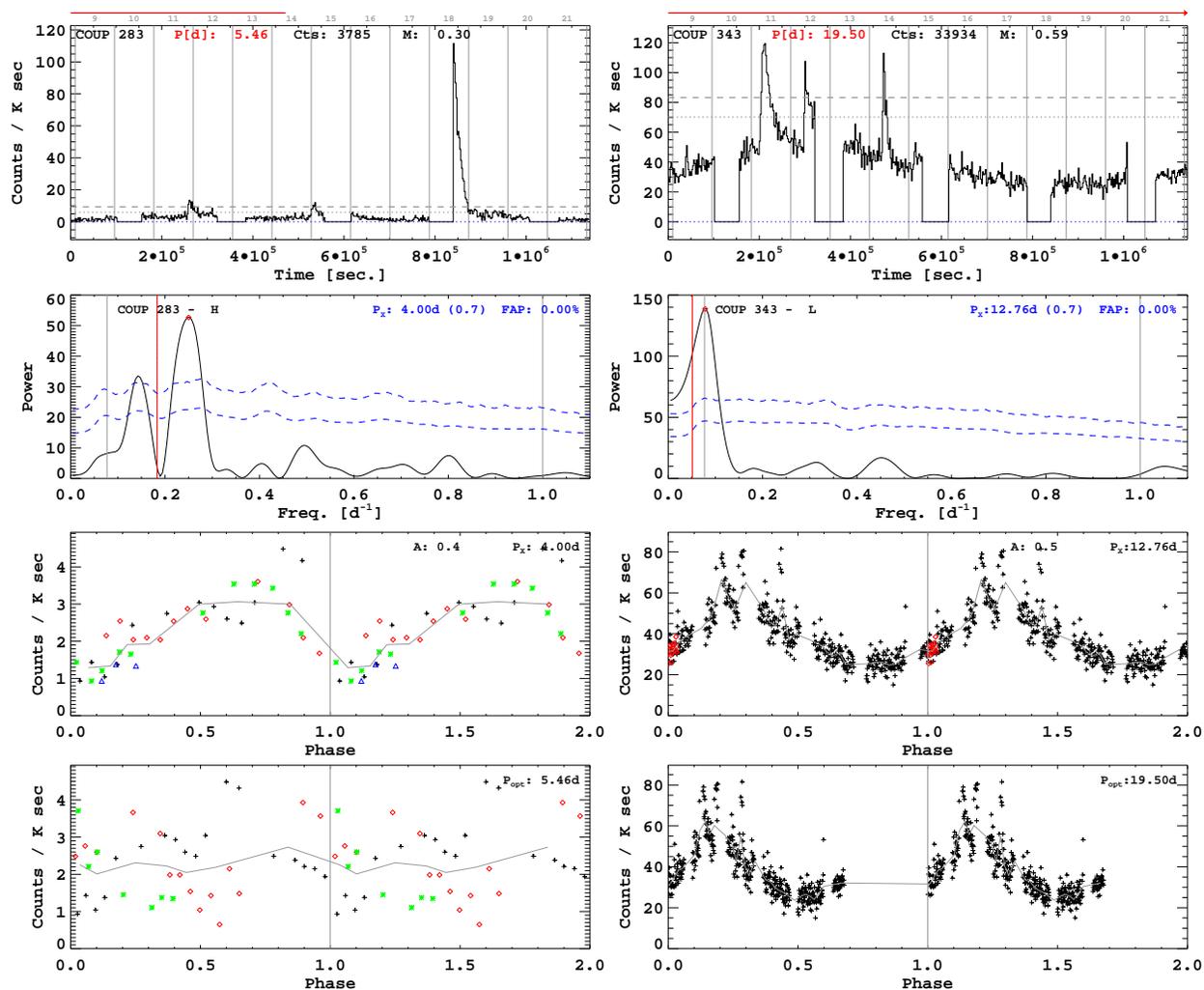}

\caption{Same as figure \ref{fig:exLC0} for two sources for which the
most likely X-ray periods is apparently unrelated to the optical one. 
\label{fig:exLC3}}

\end{figure*}

Figures \ref{fig:exLC0}-\ref{fig:exLC3} show results in four panels for
six illustrative COUP sources, three of which belong to the ``$N\cdot
L\cdot H$'' sample. We show: the original unfiltered lightcurves, the
LNP periodograms for the best filtering strategy (i.e. lowest FAP), and
lightcurves folded with both the best X-ray period and the previously
known optical period. Similar figures for the 34 $N\cdot L\cdot H$
sources in table \ref{tab:NLH} are presented in the electronic edition
of the Journal (Figure \ref{fig:exLC6}).

\subsection{Statistical comparison of $P_{opt}$ vs. $P_X$}
\label{sect:res_stat}

\begin{figure}[!t!]
\centering
 \includegraphics[width=8.0cm,clip=true]{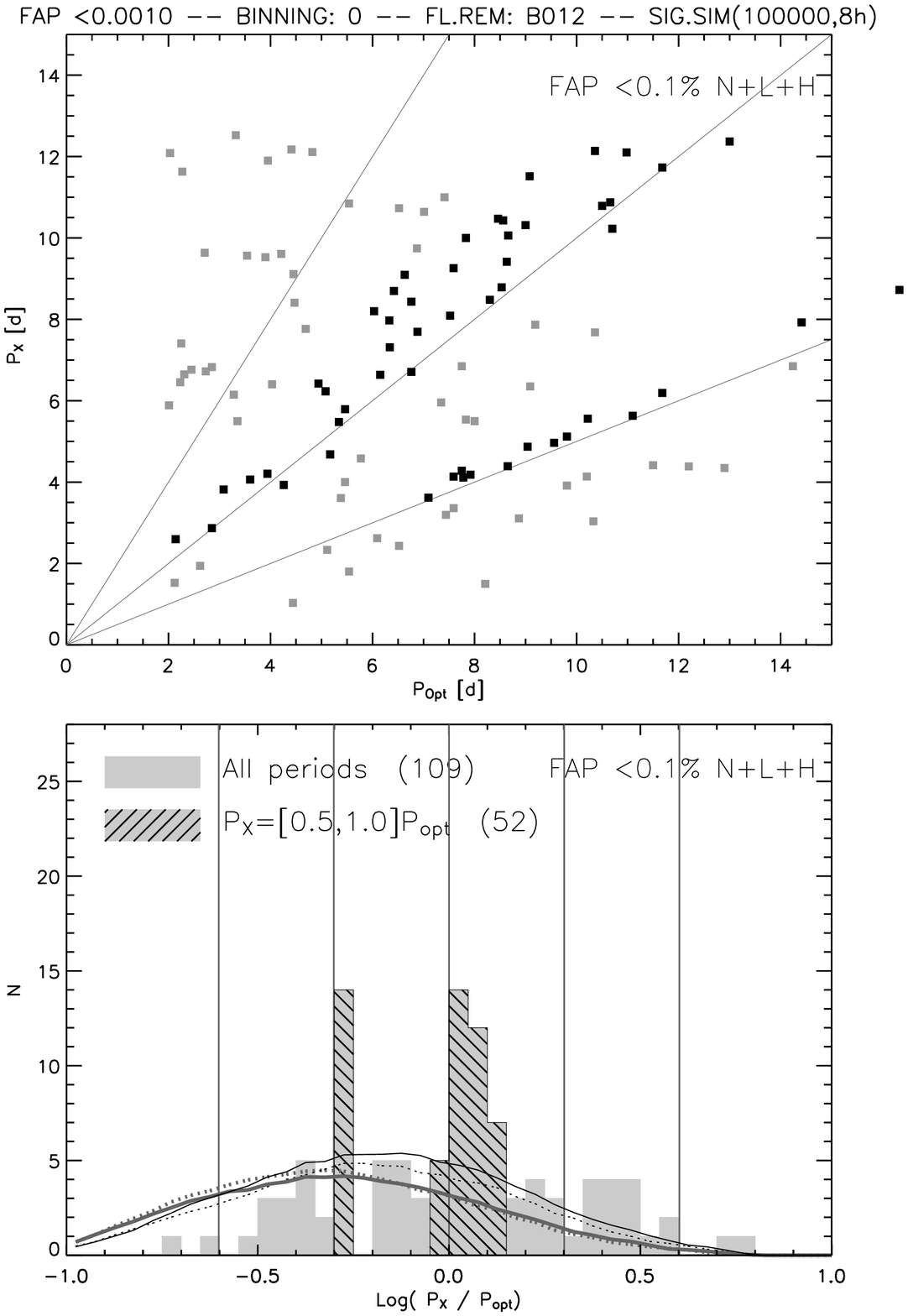}\\
 \includegraphics[width=8.0cm,clip=true]{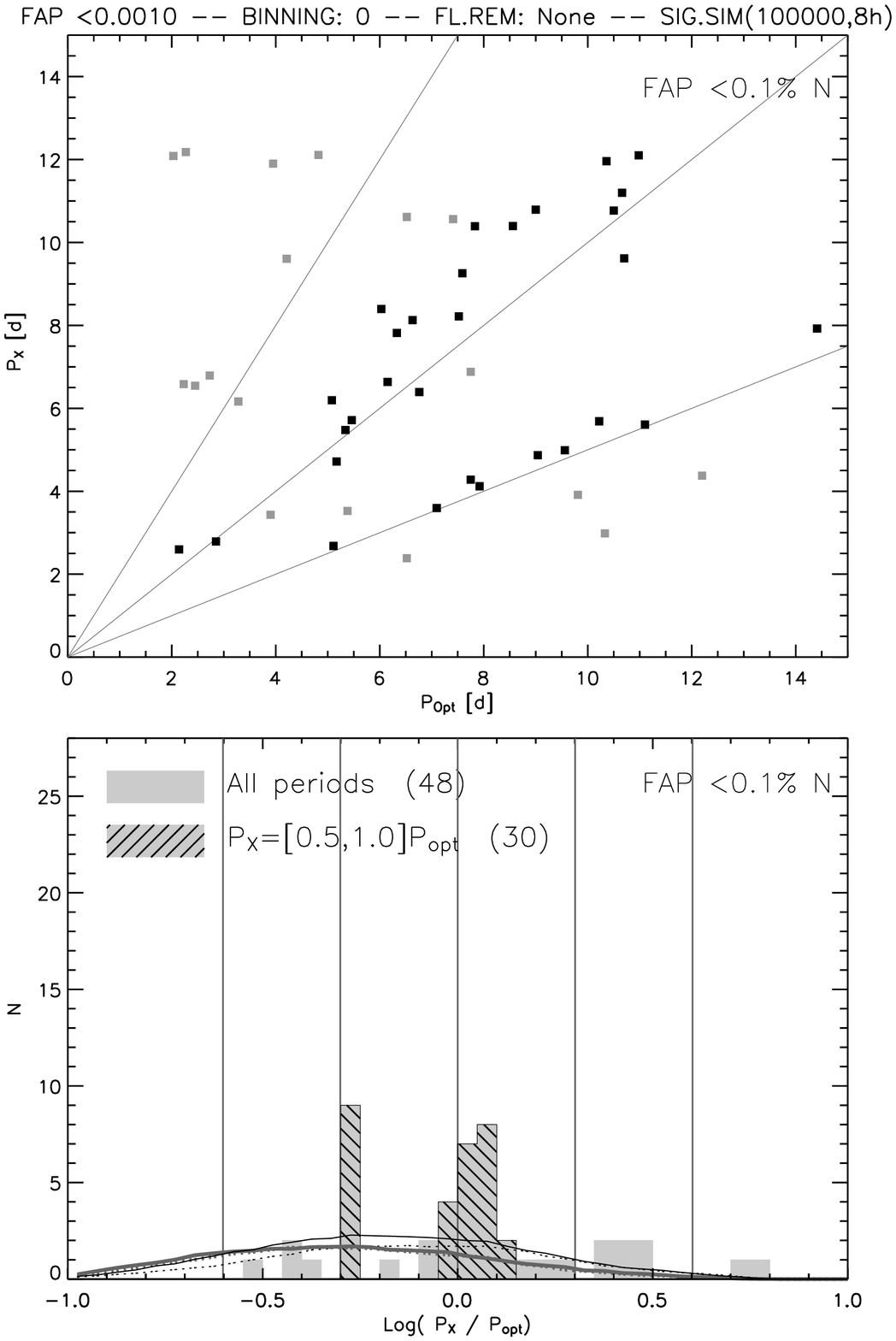}\\
 \includegraphics[width=8.0cm,clip=true]{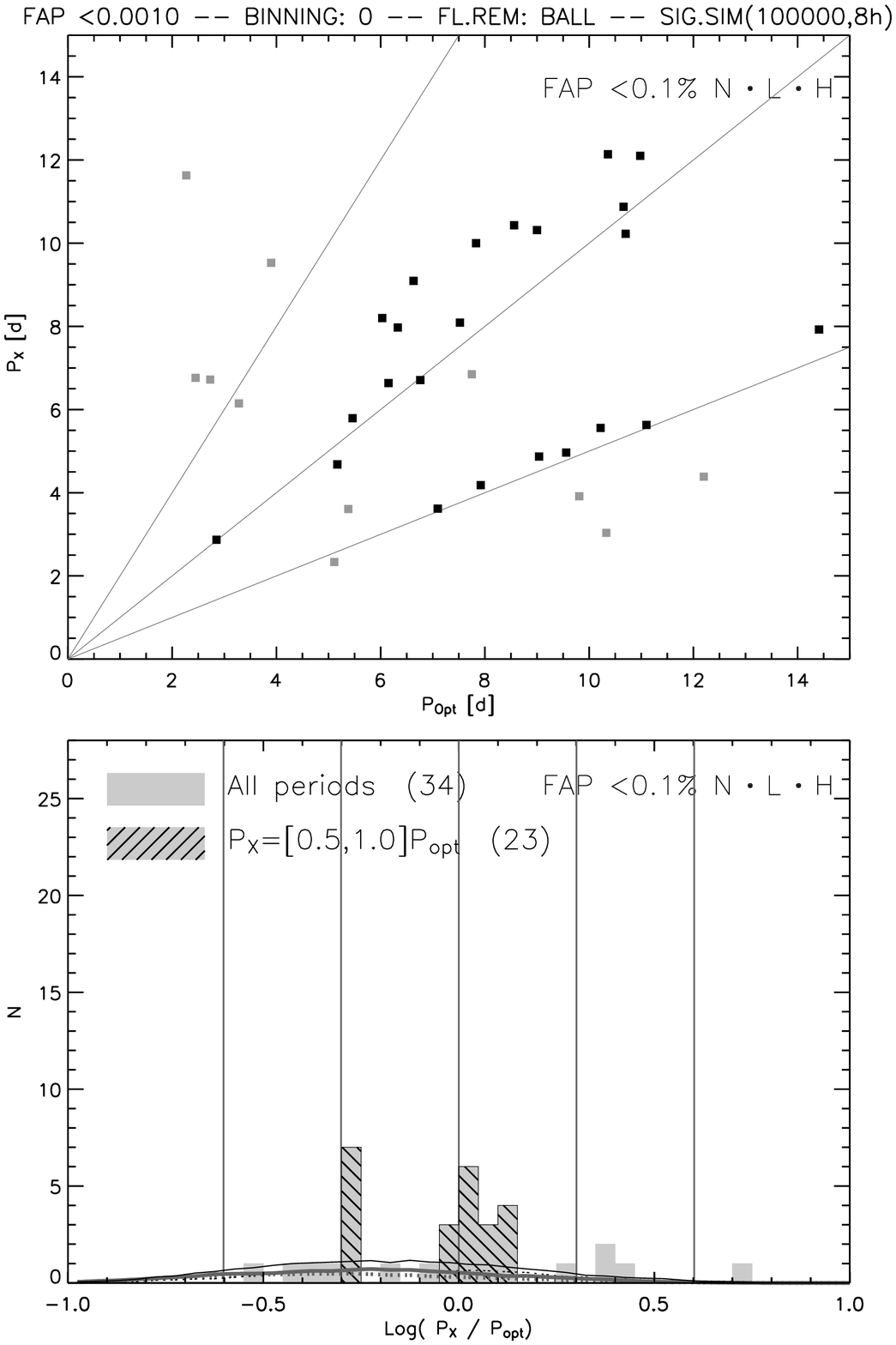}\\

 \caption{Scatter plot of the optical vs. X-ray periods. The three
panels refer to three different source samples: (Top) ``$N+L+H$'' with
FAP$<$0.1\% in at least one of the filtered lightcurves; (middle)
``$N$'' with FAP$<$0.1\% in the non-filtered lightcurve; (Bottom)
``$N\cdot L\cdot H$'' with FAP$<$0.1\% in all three filters. Darker dots
indicate sources with $P_X\sim P_{opt}$ or $P_X\sim \onehalf P_{opt}$,
i.e. those lying in the two main peaks in the histograms in Figure
\ref{fig:ratioSig}; lines indicate loci where
$P_X=[\onehalf,1,2]P_{opt}$. \label{fig:PxPo}}

\end{figure}

The three panels in Figure \ref{fig:PxPo} compare the optical and X-ray
periods for the three different subsamples discussed in the previous
section: the 109 $N+L+H$ sources showing FAP$<0.1$\% in at least one
filtered dataset; the 48 $N$ sources showing periodicity in the
unfiltered dataset; and the 34 $N\cdot L\cdot H$ sources. For the least
reliable $N+L+H$ sample, no clear relationship between the X-ray and
optical periods is seen, but for the high confidence $N\cdot L\cdot H$
sample, most of sources align close to the $P_X=P_{opt}$ and the
$P_X=0.5 \times P_{opt}$ loci.

To better define this effect, the distribution of ratios between X-ray
and optically determined periods is investigated in Figure
\ref{fig:ratioSig}. Vertical lines indicate ratios of 1/4, 1/2, 1, 2
and 4. In all samples, we see a main peak corresponding to roughly
equal X-ray and optical periods and a smaller but narrower peak
corresponding to $P_{X} \sim 0.5 \times P_{opt}$. We interpret these
plots as evidence that a substantial fraction of the X-ray periods are
indeed related to the optically determined rotation periods, even
though one is often an harmonic of the other. To strengthen this
conclusion, we show in Figure \ref{fig:ratioOth} the distribution of
$P_{X}/P_{opt}$ for X-ray sources that did not pass our significance
test, i.e. those for which FAP$>$0.1\% in all three `N', `L' and `H'
analyses (top), `N' (center), and any of `N', `L' or `H' (bottom).
These distributions show little or no relation between optical and
X-ray modulation, giving confidence that the FAP $>$ 0.1\% criterium is
indeed effective in preferentially selecting true X-ray modulation
periods.

To facilitate discussion, we distinguish three types of COUP sources:
class ``1'' when the X-ray and optical periods are nearly equal,
$-0.05< \log P_X/P_{opt}<0.15$; class ``$\onehalf$'' when when the
X-ray period is half of the optical period, $-0.3< \log
P_X/P_{opt}<-0.25$; and class ``$\neq$'' for all other stars. The light
curves and periodograms in Figures \ref{fig:exLC0}-\ref{fig:exLC3} show
two examples of each class.  These classifications are listed in the
last column of Table \ref{tab:NLH}. Out of the 109 sources in sample
{\em N+L+H},  38, 14, and 57  belong to classes ``1'', ``$\onehalf$''
and ``$\neq$'' respectively. Out of the 34 sources in sample $\rm
N\cdot L\cdot H$, 16, 7 and 11 fall in the same classes\footnote{In
some LNP periodograms of ``$\neq$'' sources, secondary peaks
corresponding to $\sim P_{opt}$ or to $\sim 0.5 P_{opt}$ are present.
If we were to consider these secondary peaks as true modulation
periods, seven ``$\neq$'' sources in the {\em N+L+H} sample and one in
the $N\cdot L\cdot H$ sample would be upgraded to class ``1''.}.

In addition to the expected finding of X-ray periods similar to the
optical periods, we thus unexpectedly find X-ray periods half as long
as established optical periods as well as a number of X-ray periods
apparently unrelated to the stellar rotation. Moreover, within class
``1'' and ``$\onehalf$'' sources, there is also a possible systematic
shift such that $P_X$ values are longer than $P_{opt}$ or $0.5\times
P_{opt}$ (Figure \ref{fig:PxPo})\footnote{We estimate the significance
of the shifts, computing the mean $\log P_X/P_{opt}$ and its
uncertainty in symmetrical intervals centered on 0 and $\log 0.5$. The
shifts are most significant in the ``N'' sample: 3.2$\sigma$ for the
$P_X/P_{opt}\sim 1$ peak and 4.8$\sigma$ for the $P_X/P_{opt}\sim 0.5$
peak}.  It remains to be seen however whether these unexpected features
in the $P_X/P_{opt}$ ratio histogram are spurious effects due to the
characteristic of the data and/or the analysis method we have applied.

\subsection{Verification of the method through simulations}
\label{sect:fsim}

Due to the complexity of the COUP lightcurves and of our analysis
procedures,  in order to better understand our results, we have
performed extensive Monte Carlo simulations of modulated and
unmodulated lightcurves that try to reproduce the essential
characteristics of our data.  Construction of these simulated
lightcurves, which include the complications of gaps in the $Chandra$
exposure and flares in the COUP stars, are described in Appendix
\ref{app:simul}. We then applied our full analysis method described in
\S \ref{sect:dataana} to these simulated lightcurves. Here we summarize
the main findings.

The dashed lines in Figures \ref{fig:ratioSig} and \ref{fig:ratioOth}
indicate the expected outcome of our period analysis on simulated
lightcurves that have no intrinsic periodicity, for reasonable choices
of model input parameters. In each panel, different lines refer to
different values of the parameters that determine the flare frequency
distribution, the lightcurve statistics and the decay time of flares,
all chosen so as to yield lightcurves that are reasonably similar to
the observed ones (specifically, in particular, $\alpha=2.50$,
$NF=[500,1000]$, $\tau_{fl}=[5,8]$hr; see Appendix \ref{app:simul}). In
all cases, these $P_x/P_{opt}$ distributions do not show any sharp peak
around 1.0 and 0.5 as we observe in the COUP data. We conclude that the
presence of these peaks must be due to intrinsic source modulation. The
fact that they occur at values of $P_X/P_{opt}\sim 1.0$ and 0.5, and
not at any other values, is a further support that the X-ray
modulations are real and related to stellar rotation.

In another important series of simulations we introduced periodic
modulations into the simulated lightcurves. These simulation show that
the observed spread in $\log P_X/P_{opt}$ among class ``1'' sources can
be attributed to uncertainties in the analysis procedure. Moreover, all
simulation series show that a fraction of sources show FAP$<$0.1\%
periods  randomly distributed between $0.2 < P_X/P_{opt} < 5$. This can
explain the observed sources with widely distributed $P_X$ values in
Figures \ref{fig:PxPo} and \ref{fig:ratioSig}. Our simulations do not
however reproduce the subtle effect that $P_X$ values tend to be
slightly longer than $P_{opt}$ in class ``1'' sources. Even more
importantly they also fail to reproduce the second peak around
$P_X/P_{opt}=0.5$. 

The fact that our simulations retrieve some of the fundamental
characteristics of the real data after our complicated data analysis
procedures lends strong support to the basic result that the class
``1'' and ``$\onehalf$'' COUP sources are truly periodic with periods
intrinsically related to the optical rotational periods.

\subsection{Physical properties of modulated stars \label{sect:physprop}}

We now return to the physical properties of the stars listed in Table
\ref{tab:sample} to investigate possible relationships to the stars
showing rotational modulation.  The ``modulated'' subsample showing
rotational periodicities are the 23 class ``1'' and ''$\onehalf$''
stars in the $N \cdot L\cdot H$ sample\footnote{Note that we will not
consider sources in class ``$\neq$''. The simulations described in \S
\ref{sect:fsim} indicate that most are likely to be spurious period
detections.}. We compare these to the full sample of COUP sources
searched for periodicities; i.e., the 233 ``searched'' sources with
known $P_{rot}$, $P_{rot}>2$d and with more than 100 detected counts.
Moreover, in order to understand possible biases in this latter sample,
we will compare the ``searched'' sample to the ``global''  sample of
COUP sources with more than 100 COUP counts.

\begin{figure}[!h]
\centering
  \includegraphics[width=15cm,clip=true]{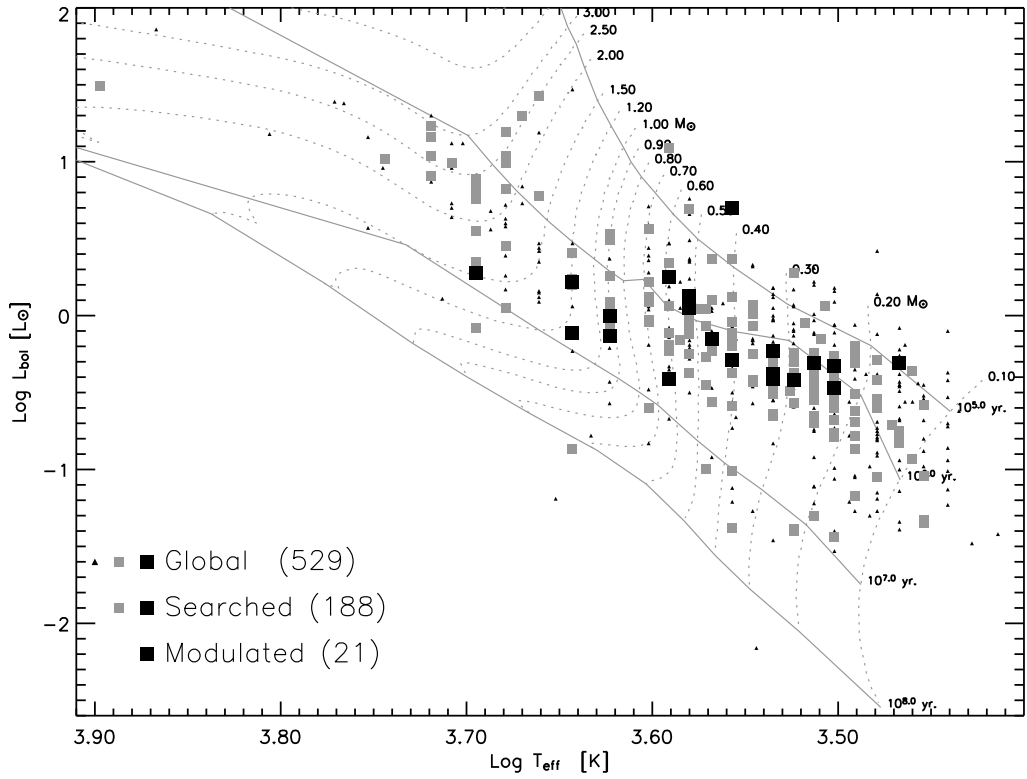}

\caption{HR diagram for all COUP sources with $>$100 ACIS counts
(``global''), for the sample for which we performed the period analysis
(``searched'') and for the subsample of this latter for which we found
X-ray periods with FAP$<$0.1\% (``modulated''). \label{fig:HR}}

\end{figure}

\begin{figure}[!h]
\centering
  \includegraphics[width=15.0cm,clip=true]{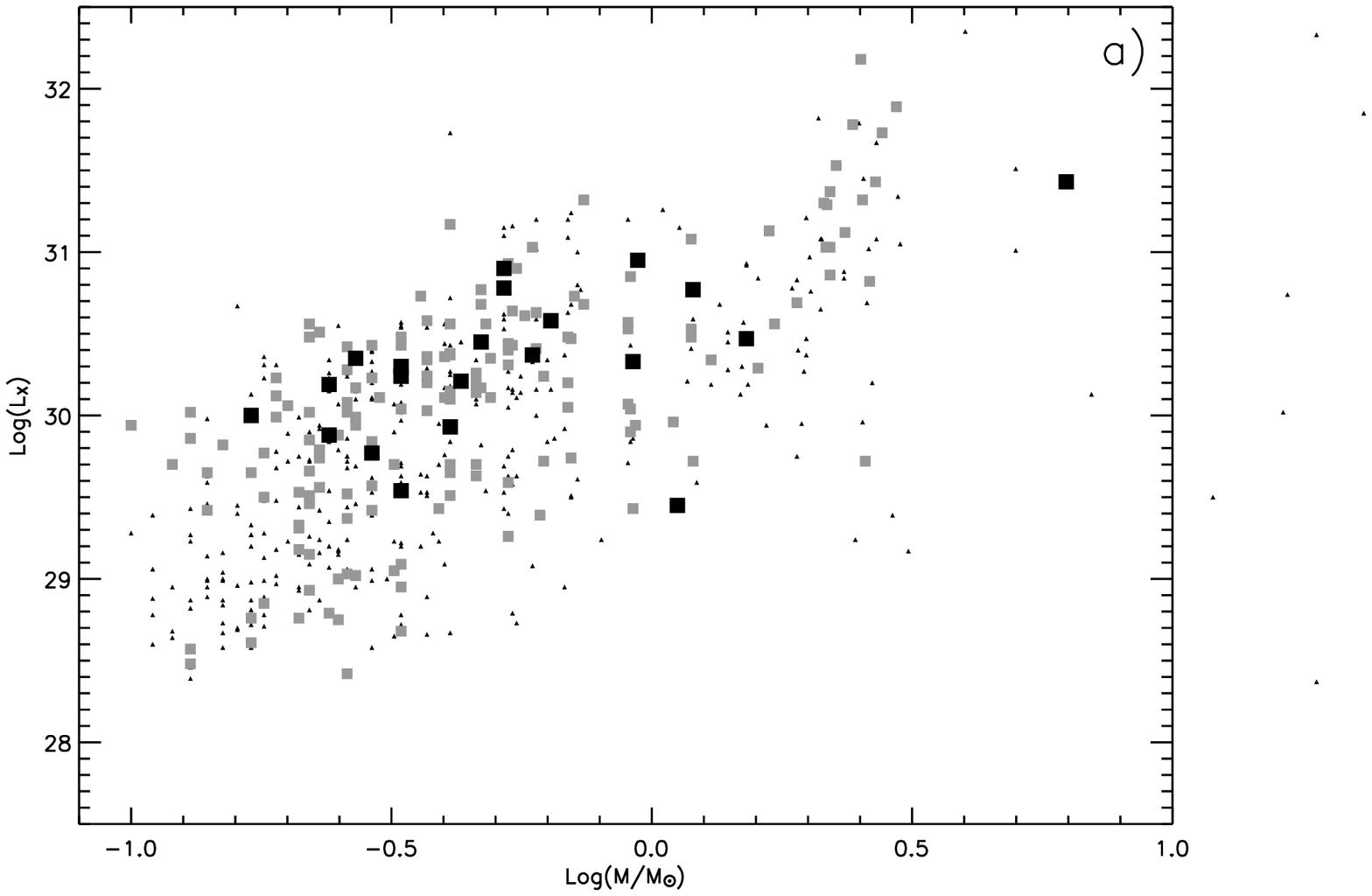}\\
  \includegraphics[width=15.0cm,clip=true]{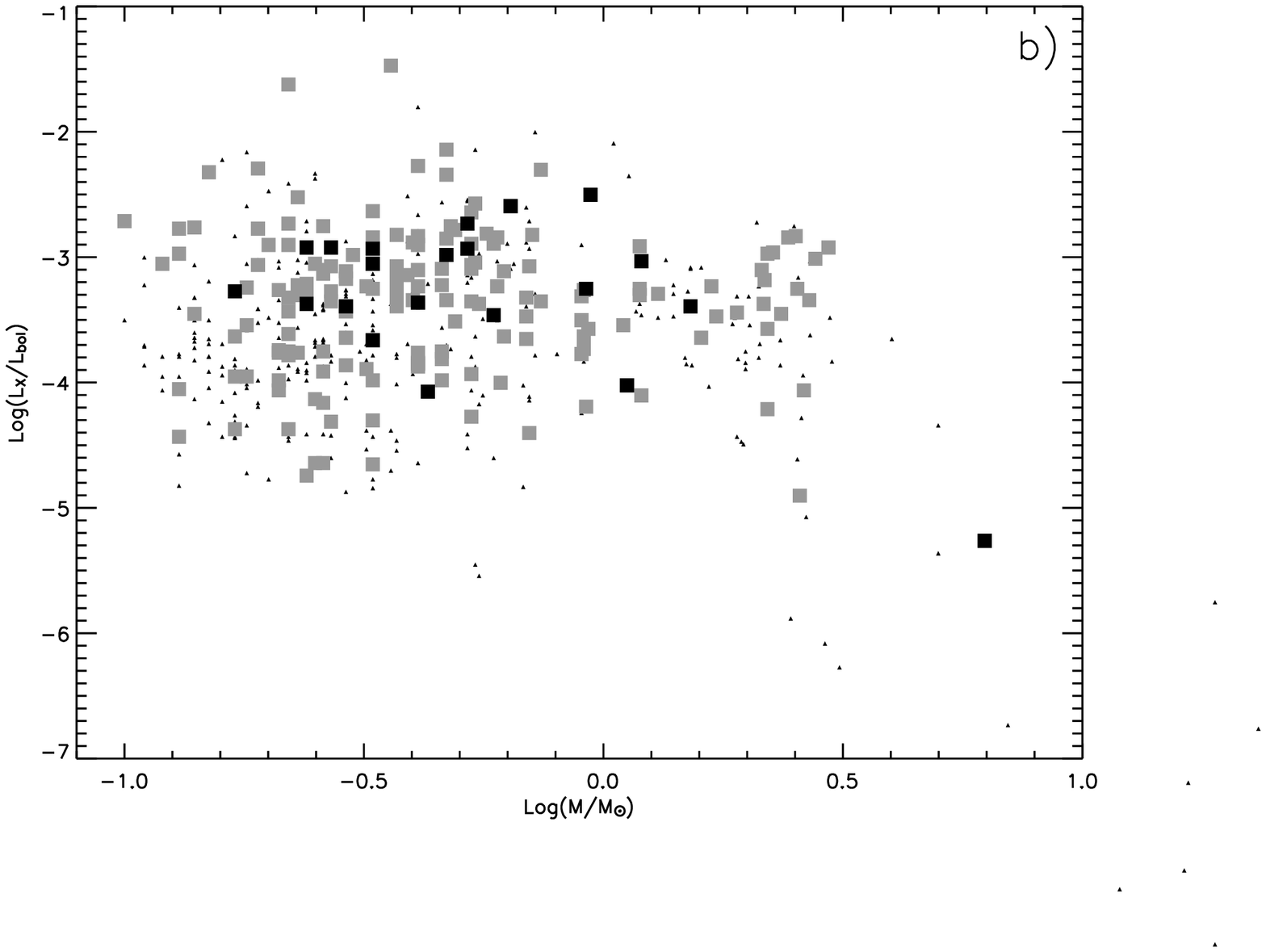}

\caption{Log(Mass) vs. $Log(L_{\rm X})$ and $Log(L_{\rm X}/L_{\rm
bol})$ for all COUP sources with mass estimates. Symbols as in Figure
\ref{fig:HR}. \label{fig:Lx_M}}

\end{figure}

Figure \ref{fig:HR} shows the Hertzsprung-Russell diagram for these
three samples.  A comparison of the mass and age distributions of the
``searched'' and ``global'' samples indicate that the former is
preferentially deficient in very low mass stars and also in the
youngest population. This may be understood as a selection effect:
very-low-mass stars are both less likely to have detected optical
modulations and have $>$100 COUP counts than more massive stars (Figure
\ref{fig:Lx_M}a). Moreover measurement of $P_{rot}$ through optical
modulation may be impeded at the earliest stellar evolutionary stages
due to high stochastic variability related to mass accretion. Compared
to the ``searched'' sample, stars in the ``modulated'' samples appear
to occupy the same area in the HR diagram and to span the full range of
stellar masses and ages. This is confirmed by two-sided
Kolmogorov-Smirnov (KS) tests that do not allow rejection of the null
hypothesis that the distributions of masses and ages are drawn from the
same parent population.

Figure \ref{fig:Lx_M}a,b show the run of $L_{\rm X}$ and $L_{\rm
X}/L_{\rm bol}$ with stellar mass for the same samples. We observe
that, respect to the ``global'' sample, stars in the ``searched''
sample, i.e.  with optical rotation periods, have larger than average
$L_X$ and $L_X/L_{bol}$ and this is confirmed with high significance by
a statistical comparison of the distributions of log$L_X$ and
log$L_X/L_{bol}$. This bias was noticed before \citep{fla03a,stas04}
and might be due to a positive correlation between coronal activity and
presence of stellar spots responsible for the modulation in the optical
band. We note that the bias appears to be stronger in $L_X/L_{bol}$;
this could result because, given a fixed photometric precision, the
probability of detecting an optical modulation depends on the contrast
between spot coverage (likely related to $L_X$) and average optical
luminosity. Within the ``searched'' sample there is a tendency to find
more X-ray modulated stars at high $L_X$ values or, equivalently, high
count-rates. For example, within the 36 stars in the ``searched''
sample with $30.5<\log L_X<31$ ergs/s, about 20\% (7 stars) are
modulated while the same fraction for the 45 stars with $\log L_X<29.5$
ergs/s is only 2\% (1 star). This effect is likely a bias due to
photons statistics: the more photons in our light curves the more
easily we detect periodicity. The $L_X/L_{bol}$ distributions of
``searched'' and ``modulated'' stars are instead statistically
indistinguishable. Due to the bias in the ``searched'' sample however
most of the ``modulated'' stars have high values of $L_{\rm X}/L_{\rm
bol}\sim10^{-3}$.

We investigated possible dependences of the periodicity detection
fraction from an indicator of mass accretion, the equivalent width of
the CaII ($\rm \lambda=8542\AA$) line, and from an indicator of disk
presence, the $\Delta(I-K)$ excess, both reported by \cite{H97}. Any
such effect would indicate a difference in the spatial distribution of
X-ray emitting plasma in the surroundings CTTS and WTTS. No striking
trend in the modulation detection fraction was found. However, with
both indicators, we found a hint (significance $\sim1$\%) that stars
with $P_X/P_{opt}\sim \onehalf$ preferentially are accreting and have
disks.

Finally we have searched for correlations of the relative
modulation amplitudes with mass, CaII, $\Delta(I-K)$, $L_X$,
$L_X/L_{bol}$, $P_X$, $P_X/P_{opt}$, without finding any.

\section{Discussion: The geometry of PMS X-ray emitting structures}
\label{sect:discussion}

\subsection{Summary of main findings and astrophysical background}

We have searched for rotational modulation of X-ray emission in a
sample of 233 ONC stars.  These are all stars with rotational period
larger than 2 days known from optical photometric studies and with more
than 100 counts collected in the COUP dataset. This sample spans the
wide mass and age range of ONC members ($0.1<M<3$ M$_\odot$,
$10^5<t<10^7$ yr). We reliably detect rotational modulation with
amplitudes between 20\% and 70\% in 23 stars or $\sim 10$\% of the
searched sample. For 16 sources, the X-ray emission is modulated at
periods very close to the optical rotational periods, while for the
remaining 7 cases the X-ray period is about one half of the optical
period. We further identify 86 stars where some evidence for modulation
is found.  Many, but likely not all, of these periodicities are
spurious due to the difficulties in accounting for X-ray flares and
other non-rotational variability. The 23 stars with rotation related
modulation have higher than average X-ray luminosities with respect to
stars in the searched sample, likely a bias due to the analysis. In all
other respects they share the same physical properties -- mass, age and
accretion -- of the searched sample. Based on these results and on
previous finding summarized below, in the next subsection we discuss
implications for the geometry of the X-ray emitting plasma on our young
stellar systems.

The high temperatures and rapid variations of PMS X-ray emission, amply
demonstrated in COUP lightcurves, implies that X-ray emitting plasma
must be confined by magnetic fields and heated by violent magnetic
reconnection events. In isolated, older, magnetically active stars,
these magnetic fields are thought to be similar to those that are
observed in the solar corona, mostly small-scale loops that connect
spots of different magnetic polarity on the photosphere.

But the configuration of the confining magnetic field around young PMS
stars might be more complicated, as these stars are often surrounded by
circumstellar accretion disks.  A variety of astrophysical models
developed with a variety of motivations suggest that the PMS stellar
magnetic field extends out $5-10$ stellar radii to the inner edge of
the disk and interacts strongly with disk material
\citep[e.g.][]{Konigl91, Calvet92, Collier93, shu94}. This results in
funneling some disk material in an accretion flow, launching other disk
material outward into jets and outflows, and locking the stellar
rotation to corotate with the inner disk. There have been suggestions
that, due to shear and instabilities, magnetic reconnection will occur
in these long star-disk loops, producing X-ray emitting plasma
\citep[e.g.][]{Hayashi96, shu97, mont00, Isobe03}.  However, the
detailed properties of X-ray emission, together with extensive evidence
for multipolar fields on PMS stellar surfaces, have generally supported
a more standard stellar activity model as applied to older stars such
as dMe flare and RS CVn systems \citep[see reviews by][]{Feigelson99,
Favata03, Gudel04}. Perhaps the first direct indication that the
observed X-rays may sometimes arise from large-scale star-disk
structures arises from detailed modeling of powerful flares observed
during the COUP observation \citep{fava05}.

\subsection{Implications of the X-ray rotational modulations}

The detection of modulation of X-ray emission at periods related to the
stellar rotation period in $\sim 10$\% of COUP stars has three
immediate implications:

\begin{enumerate}

\item The X-ray emitting structures are directly associated with
the stellar surface.  This precludes the rarely considered model
that both footpoints of the flaring magnetic loops reside in the
shearing circumstellar disk \citep{Romanova98}.

\item The X-ray emitting structures are not homogeneously distributed
in longitude on the stellar surfaces.  The amplitudes of the
modulations seen in the 23 periodic COUP stars range from 20\% to 70\%
(Table \ref{tab:NLH}, column 13), indicating that the hemispheric
brightness differences can reach roughly a factor of 2:1.

\item The bulk of the plasma emitting in the $Chandra$ $0.5-8$ keV band
is confined in magnetic structures that, in order to undergo eclipse
from the rotating star, may not extend at distances larger than $\sim
R_\star$ from the stellar surface. This conclusion is not entirely
inescapable if special geometries are considered.  For example,
eclipses could occur if the X-ray emission arises from the footpoints
of corotating star-disk loops when the disk is viewed nearly edge-on.
This and other similar possibilities we could think of seem unlikely to
occur in 10\% of otherwise ordinary PMS stars in the Orion Nebula
Cluster.  We thus view the detection of periodic X-ray modulations
related to the optical rotation period to be a solid indication that in
these stars the X-ray emitting structures responsible for the observed
modulation are compact with characteristic lengths $l \la R_\star$ or
$l << R_\star$.

\end{enumerate}

It seems likely that, due to the challenges of reliably detecting
periodic modulations in COUP stars, more than 10\% of the stars have
intrinsic X-ray periodic modulations.  Three types of stars will be
missed by our analysis: stars with small X-ray periodic amplitudes
compared to aperiodic variations such as Poisson noise and flares;
stars with large-scale inhomogeneous X-ray structures viewed too close
to their rotational axis for self-eclipsing to occur; and stars with
several or many small-scale inhomogeneities. In this last case, some
bright features emerge from eclipse as others enter eclipse, so the net
rotational modulation has smaller amplitude than the smaller-scale
spatial variations. This scenario of multiple X-ray hot spots near the
surface is supported by our discovery of several COUP stars with $P_X =
0.5 P_{opt}$, attributable to the eclipses and emergences of bright
areas on opposing hemispheres. We thus conclude that {\it at least}
10\% of COUP stars have large-scale structure in the longitude
distribution of the X-ray emitting loops on scales less than a stellar
radius.


Perhaps the most valuable inference that emerges from our findings
concerns the enigma of `saturation' in magnetically active stars. Our
low mass modulated stars have very high levels of $\log L_X/L_{bol}$
(Figure \ref{fig:Lx_M}b). Excluding one high mass star (COUP 1116,
$M\sim 6M_\odot$) the median value is -3.05, basically the saturation
level seen in magnetically active stars. At least for these stars, we
can exclude one of the several suggested explanations for saturation
(\S \ref{sect:intro}): it can not arise due to an uniformly filling of
the stellar surface with X-ray emitting plasma.  A surface fully
covered with X-ray loops would limit rotational modulation to a very
low level. We note that, if we assume a solar-like picture for the
X-ray emitting regions and their relation with stellar photospheres,
our finding is in agreement with the existence on these stars of
photospheric inhomogeneities similar to solar spots, as inferred from
optical rotational modulation.

\subsection{Comparison with the Sun}

It is helpful to consider the Sun, the only star for which we have
directly imaged longitudinal X-ray structures over many years. The
astronomical and astrophysical similarities between the Sun and
magnetically active stars has been well-documented \citep{Schrijver00}.
It is important to keep in mind that the solar X-ray corona is
certainly very different from that of our PMS stars.  It is rotating
much slower, has a much softer X-ray spectrum (even during flares), and
is hugely less luminous.  Latitude distributions of active regions may
differ in PMS stars, and viewing orientations will be random rather
than optimized for rotational eclipses.

\begin{figure}[!t]
\begin{center}
\includegraphics[width=16cm]{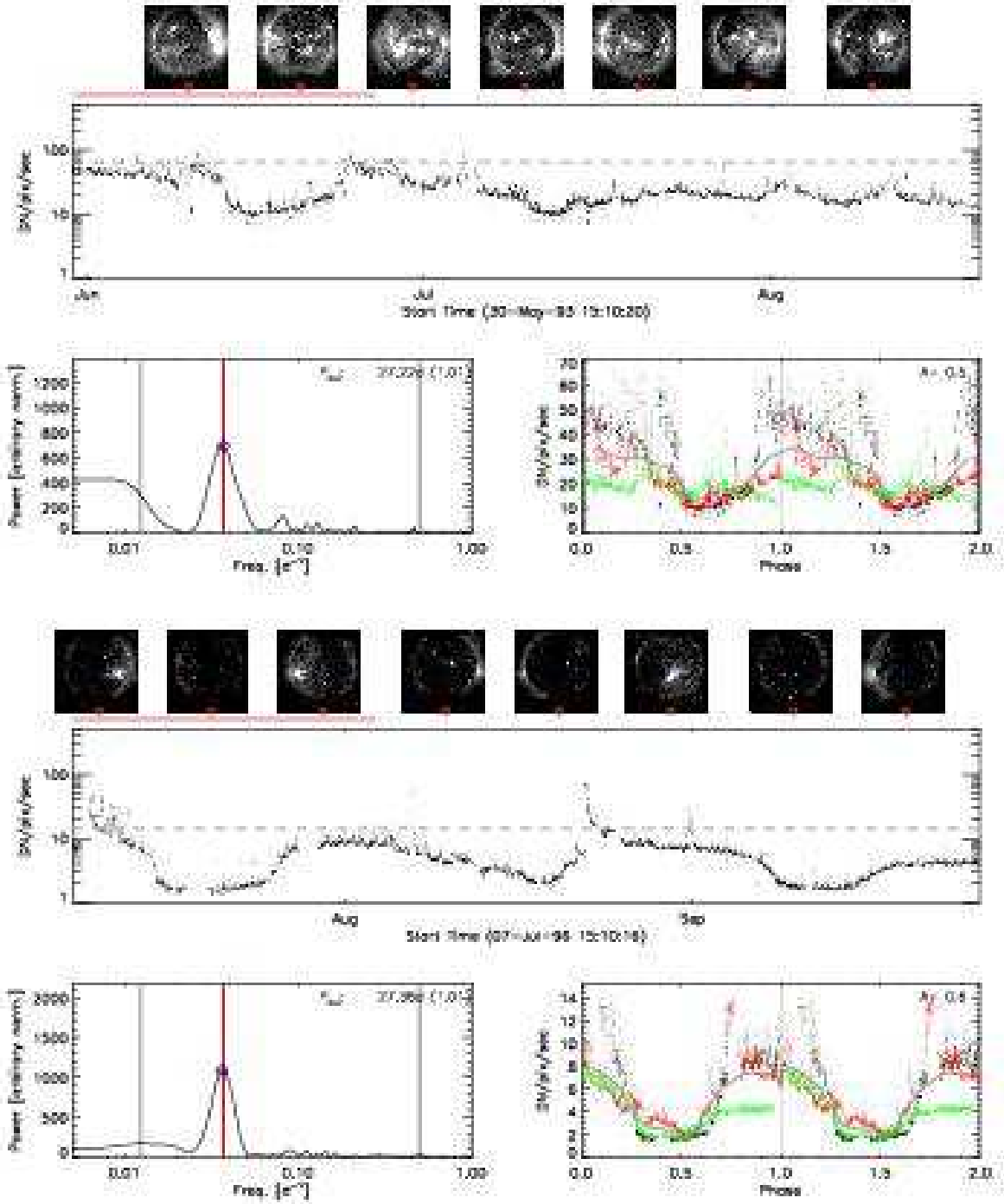}

 \caption{YOHKOH SXT (AlMg filter) light curves with relative full disk
images (at selected times), periodograms and folded light curves, in a
format similar to figure \ref{fig:exLC0}. The two sets of panels refer
to different 81 days time segments: the upper one is close to solar
maximum, the lower one to solar minimum. \label{fig:sun_lnp}}

\end{center}
\end{figure}

The YOHKOH Soft X-ray Telescope \citep{tsune91} has obtained an almost
continuous time series of disk integrated solar fluxes comprising $\sim
150,000$ measurements in its AlMg band ($\sim 0.4-4$ keV) between 10
Nov 1991 and 21 Dec 2000 (both at solar maximum) and including one
solar minimum around 1996-1997. Due to the SXT operating mode, times
during the most energetic flares are automatically excluded from the
time series. We have analyzed this dataset in a fashion similar to that
performed on the ONC stars. We first divide the data into contiguous 81
day long segments, each comprising three solar rotation periods. The
LNP analysis is then applied to each segment as described in \S
\ref{sect:dataana} searching for periods between 2 and 81 days. Figure
\ref{fig:sun_lnp} shows the typical results for four segments.  As with
COUP stars, the panels show the X-ray lightcurves, LNP periodograms,
and the lightcurves folded with the best period. On top of light curve
we show YOHKOH images of the solar corona at corresponding times. In
all the cases shown the periodograms and folded lightcurves refer to
the lightly trimmed (``L'') lightcurves.

Our procedures succeed in retrieving the solar rotation period $P
\simeq 27.3$ d in $\sim 75$\% of the temporal segments. This is a much
higher success rate than for the ONC sample. Three factors make period
finding easier for the solar case: extremely strong signals; optimal
inclination of the rotation axis; and flares (those not automatically
excluded from the time series) with durations much shorter than the
rotation period. Modulation amplitudes computed from the folded
lightcurves, range between 20\% and 60\% (median $\sim 0.35$\%),
similar to the rotationally modulated COUP stars. There might be a
tendency to have slightly lower relative amplitudes at solar maximum
rather than at minimum. The shape of the light curves vary
significantly between segments depending on the configuration of active
regions; broadly speaking they are similar to the ones we observed in
our PMS stellar sample.

From this first comparison between stellar COUP data and the solar
YOHKOH data we conclude that, although in PMS stars plasma emission
measures and temperatures are tremendously higher, the degree of
longitudinal inhomogeneity in the X-ray emitting regions are
qualitatively similar. It is thus reasonable to view both the
2-dimensional surface distribution -- dominated by one or a few active
regions -- and the 3-dimensional distribution -- dominated by emission
close to the stellar surface -- of X-ray emission in the Sun to be
similar to that in PMS stars.

\section{Summary}
\label{sect:summconc}

We have searched for rotational modulation of X-ray emission in a
sample of 233 young PMS stars in the Orion Nebula Cluster. The study
was made possible by the {\em Chandra Orion Ultradeep Project} which
obtained an 850ks long ACIS observation of the region during a time
span of 13.2 days. In order to check that the observed X-ray modulation
is related to stellar rotation, this work focused on a selected sample
of stars with know rotation period. 

Periodicity in binned light-curves was searched with the Lomb
Normalized Periodogram method. The analysis was performed both on
unfiltered lightcurve and on lightcurves that were first trimmed to
remove large flares. False alarm probabilities were estimated so as to
take into account the presence of non-rotational short-term ($<8$ hour)
variability such as flares; because of non-rotational variability with
longer timescales these FAPs remain indicative. We thus verified our
analysis technique using extensive numerical simulations.

We reliably establish the presence of X-ray rotational modulation in 23
stars, 10\% of the searched sample and the largest sample of stars of
any class in which X-ray rotational modulation has been observed to
date. For 86 additional stars in the searched sample the detection of
rotational modulation is less reliable. Within the 23 reliable
detections relative amplitudes range from 20\% to 70\% and periods
between 2 and 12 days. In 16 cases the X-ray modulation period is
similar to the stellar rotation period, while in seven cases it is
about half that value. The data suggest that X-ray periods are on
average 5-10\% longer than either $P_{opt}$ or $0.5 P_{opt}$.

Comparing the stellar properties of the modulated sample to those of
the searched sample, we find that the two sample are indistinguishable
with respect to mass, age and accretion disk properties. X-ray
modulation is however detected preferentially in bright X-ray sources.
We understand this as a selection effect. The data also suggest, albeit
with low significance, that stars with $P_X \sim 0.5\times P_{opt}$ are
preferentially active accretors, according to the CaII line EW, and
posses disks, according to the $K$ band excess.

Our main conclusions, referred to the stars that show modulation, are:

\begin{enumerate}

\item X-ray emitting plasma is inhomogeneously distributed in
longitude. A comparison with solar data suggests a similar degree of
inhomogeneity.

\item Saturation of activity, $L_X/L_{bol}\sim 10^{-3}$, is not due to
the filling of the stellar surface with active region.

\item Dominant emitting structures are likely compact with sizes
$\lesssim \rm R_\star$.

\end{enumerate}

The last conclusion may seem in contrast with that of \cite{fava05}
based on the analysis of the decay phase of the most luminous flares
observed during the COUP observation. \cite{fava05} find that a number
of large flares can be modelled only by assuming very long magnetic
loops ($5-20~R_\star$). It is not yet clear how common these large
loops structures are. It is conceivable that long-lived compact
structures are the most common and that they coexist with extended ones
that may become prominent during flares.

\begin{acknowledgements}

The authors are grateful to K. Stassun, S. Shang and S. Orlando for
useful discussion. COUP is supported by $Chandra$ Guest Observer grant
SAO GO3-4009A (E.\ Feigelson, PI). E.F, G.M. and S.S. acknowledge
financial support from the {\em Ministero dell'Istruzione
dell'Universit\'a e della Ricerca}.  E.D.F. is also supported by ACIS
Team contract NAS8-38252.

\end{acknowledgements}

\clearpage
\newpage

\appendix

\section{Simulations of COUP X-ray lightcurves}
\label{app:simul}

This Appendix describes the development of simulated X-ray lightcurves
to test the performance of our analysis techniques in datasets
suffering the realistic complexities seen in COUP ONC X-ray
lightcurves.  These include repeated high-amplitude flaring,
longer-timescale aperiodic variations, gaps in the exposure due to the
$Chandra$ orbital perigees, and Poissonian noise. Other currently
unknown sources of variability might also be present.  We focus much
attention on the effect of flares on the period detection procedures.
Simulating realistic light curves is not straightforward because it
requires the knowledge of several input quantities that are at present
not well constrained, such as the flare frequency, distributions of
flare amplitudes and durations, the relative contribution of flares and
``quiescent'' emission, if at all present.  These issues are studied
for a small sample of 1~M$_\odot$ COUP stars by \citet{wolk05}, but
have not yet been characterized for the full COUP sample.

We therefore introduce a model of the stellar X-ray emission that is
simple enough to be described with few parameters, but still yields
lightcurves that qualitatively resemble observed ones. The model
lightcurves are made exclusively out of a distribution of flares based
on the well-studied idea that even stellar ``quiescent'' emission
arises from the superposition of many microflares
\citep[][and references therein]{Kashyap02}. Forthcoming COUP
studies will tackle this issue in greater detail. Here, we
introduce this model exclusively as a tool to produce useful lightcurves.

\subsection{Construction of simulated lightcurves}

The following procedure was followed for creating unmodulated
lightcurves\footnote{Throughout these simulation we will neglect
background contributions to the light curves. This is justified by the
relative uninportance of background for the majority of COUP sources
\citep{get05}, implying that the noise properties of real
background-subtracted lightcurves are very similar to those of
background-free simulated lightcurves.}:

\begin{enumerate}
\item Generate a specified number, NF, of random flare start times,
uniformly distributed between $t_0-3\tau_f$ and $t_1$ where $t_0$,
$t_1$ are the start and end times of our COUP observation
and $\tau_{fl}$ is a specified decay timescale for flares.

\item Assign to each of the NF flares an intrinsic amplitude, $C_0$ or
total number of counts, obtained from the probability distribution
$\frac{dn}{dC_0} \propto C_0^{-\alpha}$ where $\alpha$ is a specified
powerlaw index of the flare intensity distribution.  We find (see
below) that the observed data is best reproduced by $\alpha$ larger
than two. To avoid an energy divergence of small flares we arbitrarily
adopt a lower limit $C_0 > C_{min} = 1$ of flare
intensities\footnote{Corresponding to a minimum flare energy $\sim
5\times 10^{32}$ ergs.}.

\item Compute the observed amplitude $C$ for each flare, randomly taken
from a Poisson distribution with mean $C_0$.

\item Simulate $C$ photons arrival times for each flare, assuming an
instantaneous rise and an exponential decay with a decay time $\tau_{fl}$.

\item Merge the arrival times for each flare so to generate the simulated
light curve.

\item Filter out simulated events that fall outside the Good Time
Intervals of the COUP observation (i.e. in the gaps due to the
$Chandra$ satellite orbital perigee).

\end{enumerate}

\begin{figure}[!h]   
\includegraphics[width=16cm]{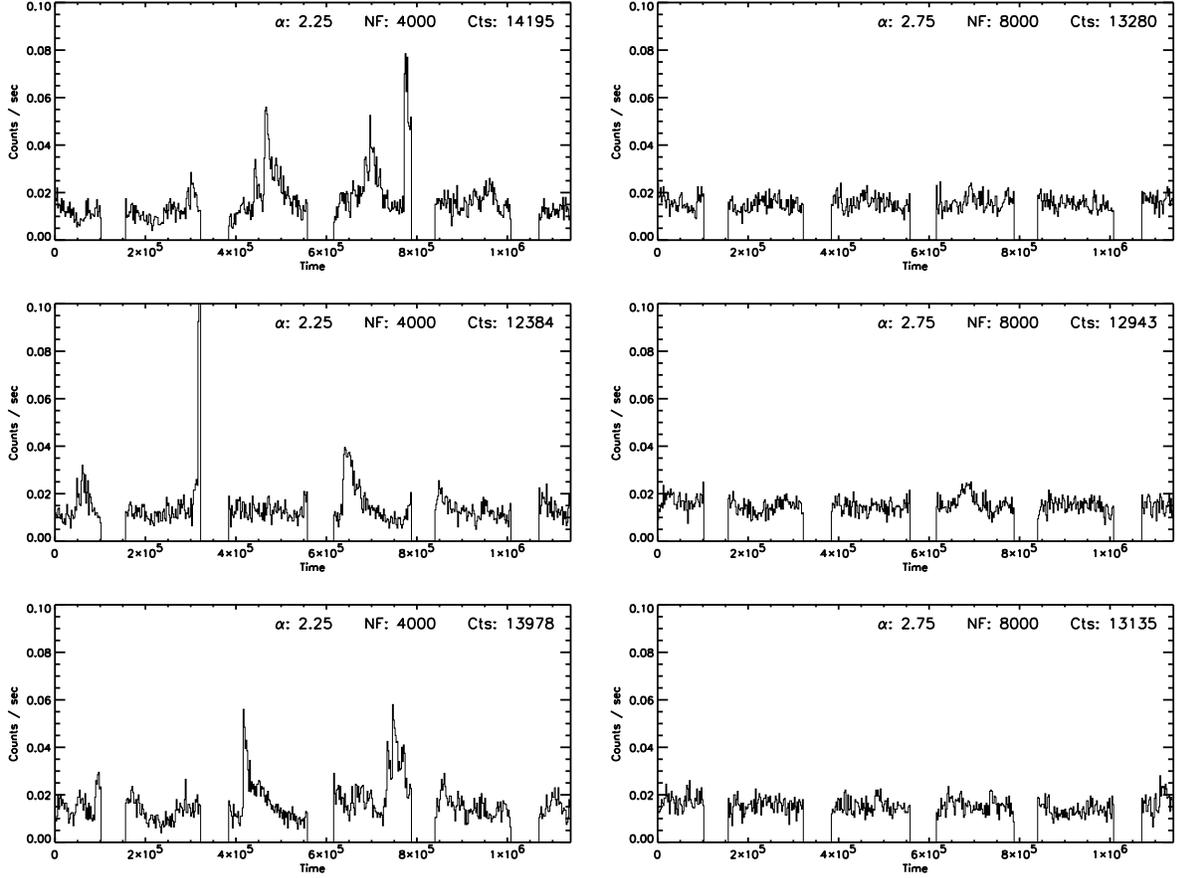}

 \caption{Example of six simulated lightcurves with no modulation,
 three for each of two values of the parameter $\alpha$ (= the slope of
 the flare intensity distribution): $\alpha=2.25$ (left) and 2.75
 (right). The number of simulated flares ($NF$), all having
 $\tau_{fl}=5h$, is chosen so to yield, in the two cases, a similar
 average number of counts. The actual number of simulated counts is
 given in the upper right corner of each panel. \label{fig:sim_lc}}

\end{figure}

Having fixed $C_{min}$, the free parameters in the model are $\alpha$,
the steepness of the flare amplitude distribution, $NF$, the number of
flares, and $\tau_{fl}$, the flare decay time. We consider two values
of $\tau_{fl}$, 5 hr and 8 hr, consistent with the typical flare
durations seen in COUP lightcurves. Setting $NF$ is equivalent to
setting the total counts for the source. We examined lightcurves
produced with values of $\alpha$ ranging between 2.0 and 3.0. The
flatter $\alpha$=2.0 distribution produces lightcurves with far too
many bright flares compared to typical COUP lightcurves, while
$\alpha$=3.0 clearly results in too few bright flares. Figure
\ref{fig:sim_lc} shows six examples of simulated light curves.

In order to simulate rotational modulation, we modify the distribution
from which flare start times are drawn (step 1) from a constant
distribution to a sinusoidal one with a given input period ($P_{in}$)
and amplitude ($Amp$). We then apply our period searching methodology
(\S \ref{sect:dataana}) to all simulated lightcurves, including
attempts to trim flares, calculate LNPs, estimate FAPs, and select
significants results according to various criteria.

We initially ran these simulations choosing the input parameters from
the following grid: $\alpha$= 2.0 to 3.0 in 0.25 steps, $NF$= 500, 1000
and 2000, Amp= 0\% to 100\% in 20\% steps. $P_{in}$ were chosen equal
to the optically determined rotation periods of the 233 stars in our
main sample, repeated five times so as to have, for each choice of the
the other parameters, 1165 simulated light curves with $P_{in}$
distributed as $P_{rot}$. The value of $\alpha$ was constrained using
two simple strategies to compare real and simulated lightcurves with a
similar number of counts: visual examination of simulated and real
lightcurves, and comparison of the standard deviation of the count
rates in the binned lightcurves. Both of these qualitative comparisons
indicate that $\alpha$ between 2.25 and 2.5 gives light curves that
most resemble the observed ones. With these choices of $\alpha$, the
$NF$ (number of flares) that reproduces the median observed number of
counts in our ``searched'' sample ($\sim 2300$) is between 500 and
1000.  Simulations using these parameters, with and without periodic
modulations applied,  were then used to study the performance of our
period finding method.

\subsection{Results for simulated unmodulated COUP sources}

Simulations with $Amp=0.0$ are valuable to testing the false alarm
probabilities which may lead to spurious period detections. Figure
\ref{fig:simA00} shows, in the same format of Figure \ref{fig:ratioSig}
with actual COUP data, the distributions of $\log P_{out}/P_{in}$, for
the three selections of ``significant'' periods (``$N+L+H$'', ``$N$'', 
``$N\cdot L\cdot H$'') and for our chosen ranges of model input
parameters. The histograms are normalized to yield the expected
distributions for our ``searched'' sample. Since the simulated
lightcurves are not modulated, there is no relation between input and
output period. The shape of the simulated distributions therefore
solely reflects the distribution of {\em spurious} detected X-ray
periods with respect to the input optical stellar rotation periods. The
distributions show that using a realistic model of the X-ray emission
from ONC stars:

\begin{enumerate}

\item Detection of spurious periods due to our failure to properly account
and/or remove flares, cannot explain the observed sharp peaks in the
observed $P_X/P_{opt}$ histogram.

\item Flares can account for most or all of the X-ray periods that do not
fall within these peaks and that are therefore unrelated to optical
periods. It is also apparent that the number of spurious periods found
in these unmodulated lightcurves is largest when accepting periods
that have $FAP<0.1$\% in any of the three analyzed light curves (the
``$N+L+H$'' sample) and lowest when accepting only periods with
$FAP<0.1$\% in all three analyses (the  ``$N\cdot L\cdot H$'' sample).

\end{enumerate}

\subsection{Results for simulated modulated COUP sources}

\begin{figure}[!h]
\centering
 \includegraphics[width=8.0cm,clip=true]{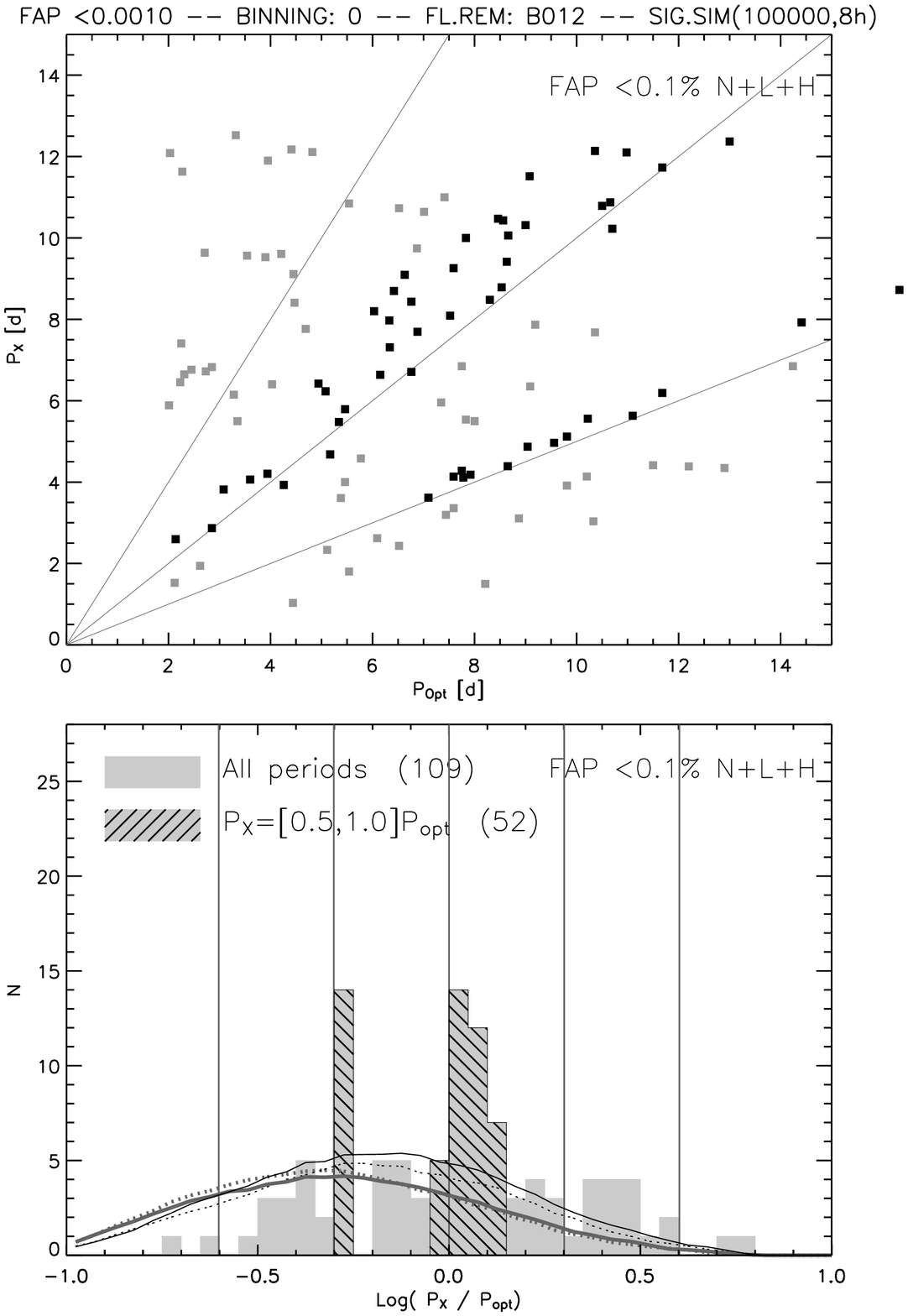}\\
 \includegraphics[width=8.0cm,clip=true]{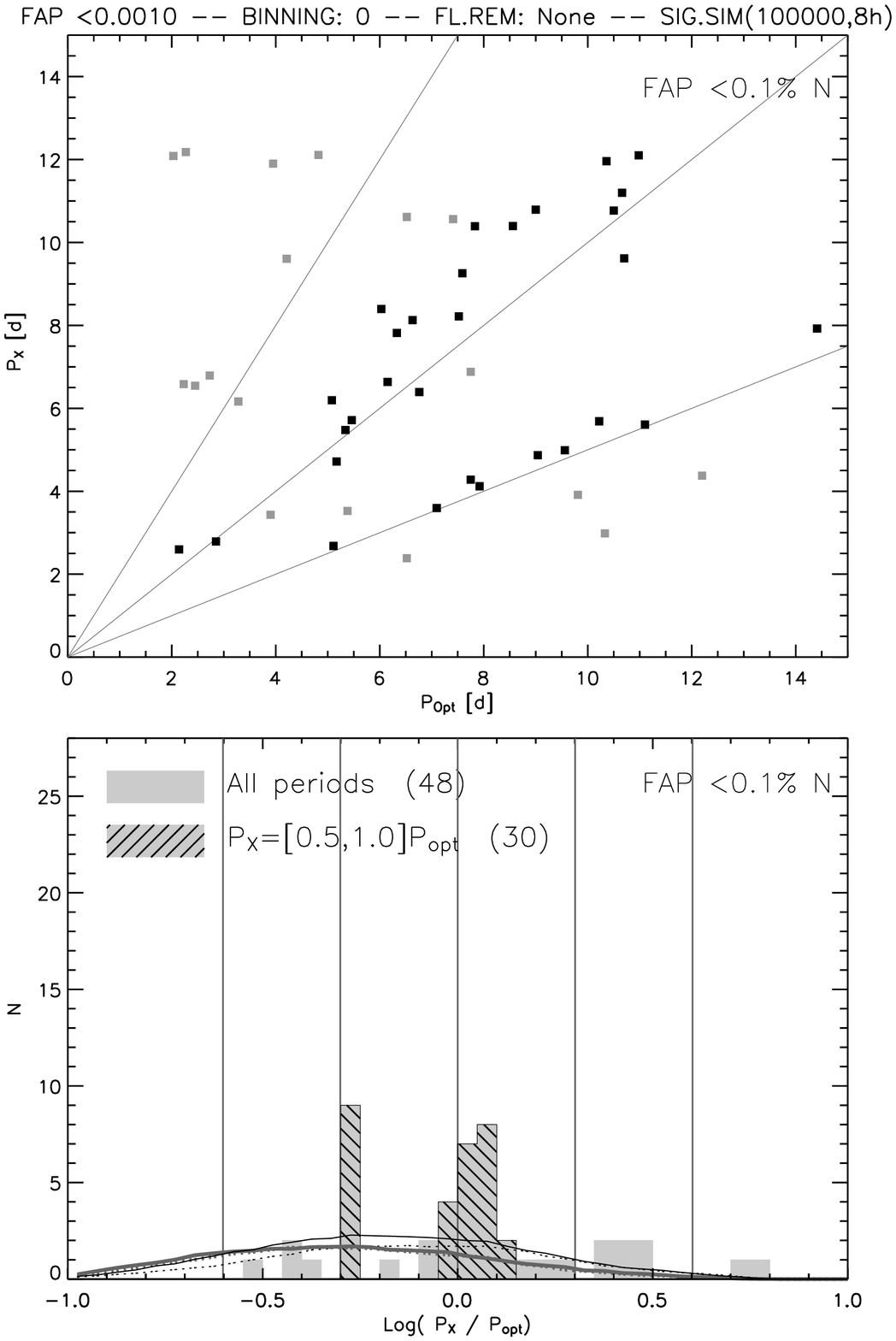}\\
 \includegraphics[width=8.0cm,clip=true]{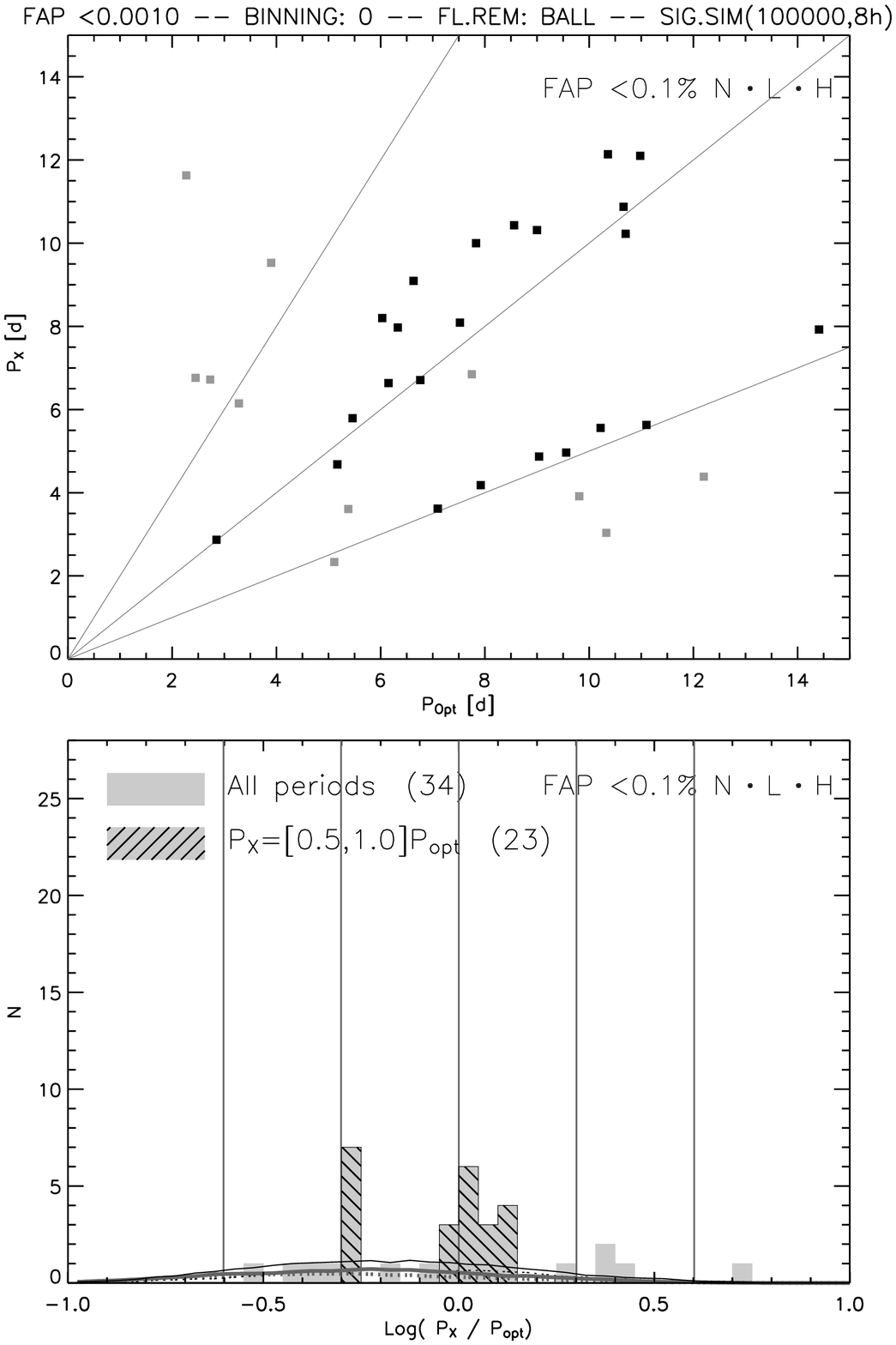}\\

 \caption{Distribution of the logarithm of the ratio between X-ray and
 optically determined periods for X-ray periods with FAP $<0.1$\% and
 three different choices of light curve filtering strategy as in Figure
 \ref{fig:PxPo}. Also shown (curves) are distributions resulting from
 the simulation of unmodulated flaring lightcurves (see Figure
 \ref{fig:simA00}). Vertical lines indicate ratios of 1/4, 1/2, 1, 2
 and 4.\label{fig:ratioSig}}

\end{figure}

\begin{figure}[!h]
\centering
 \includegraphics[width=8.0cm,clip=true]{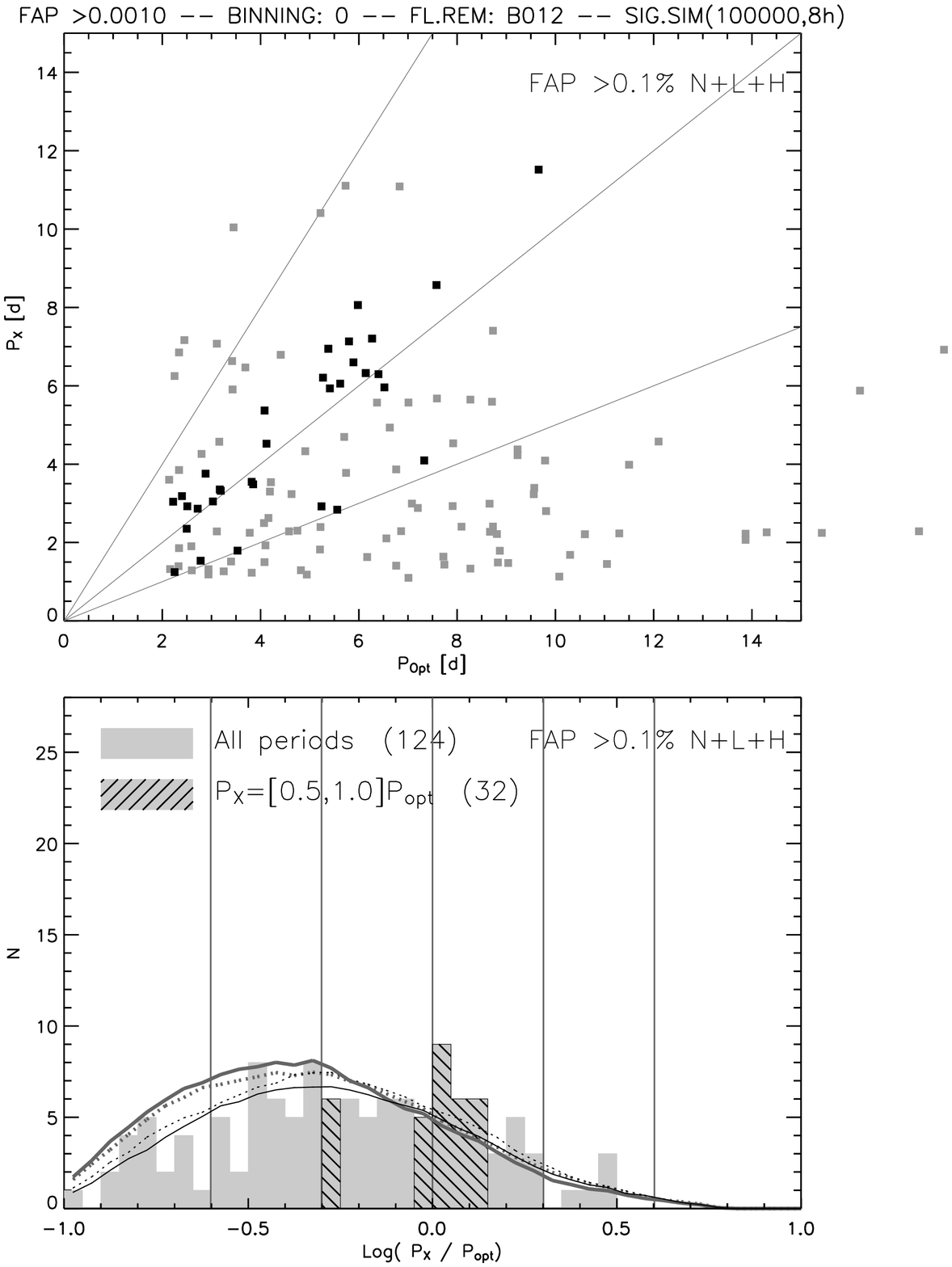}\\
 \includegraphics[width=8.0cm,clip=true]{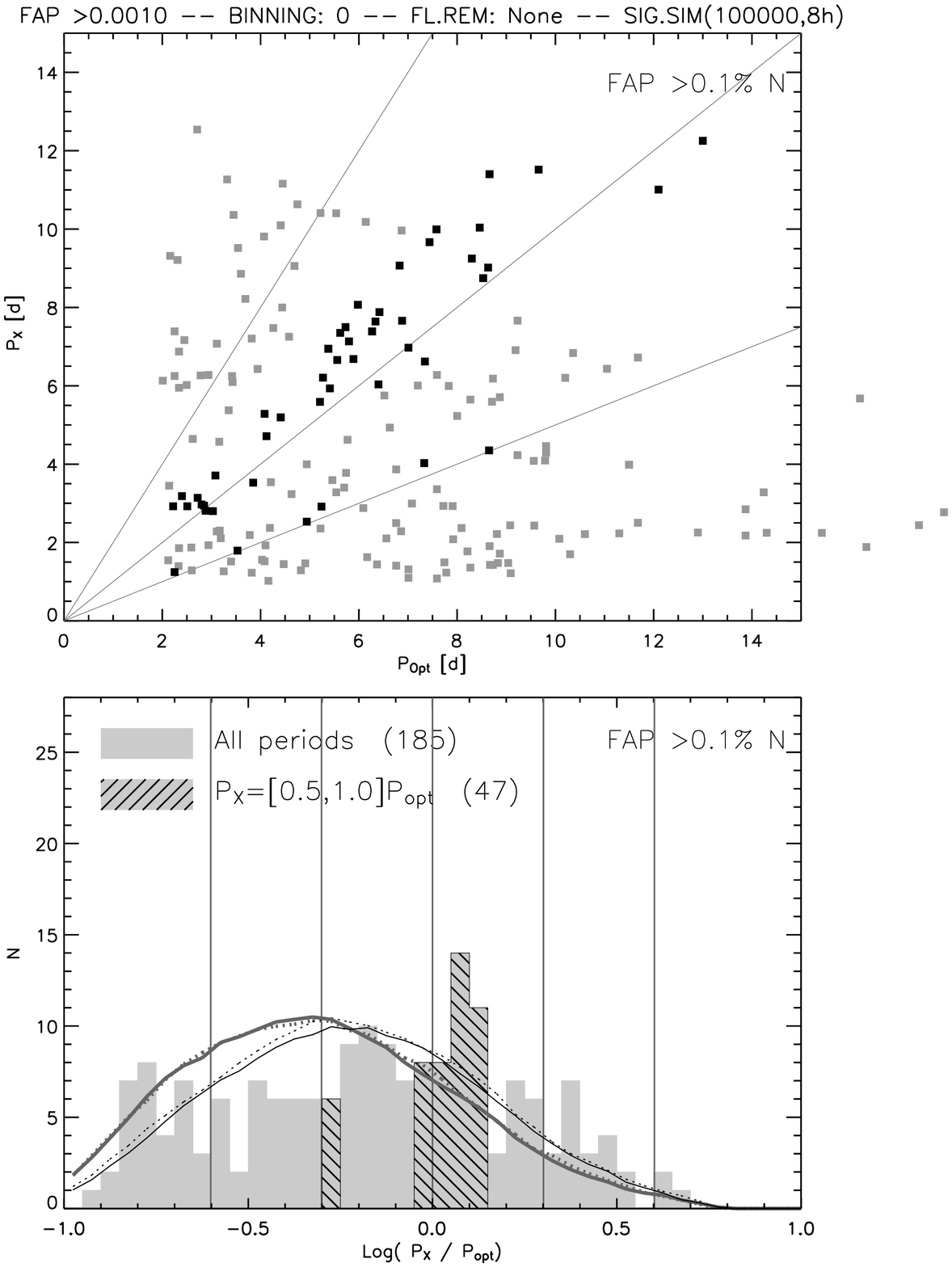}\\
 \includegraphics[width=8.0cm,clip=true]{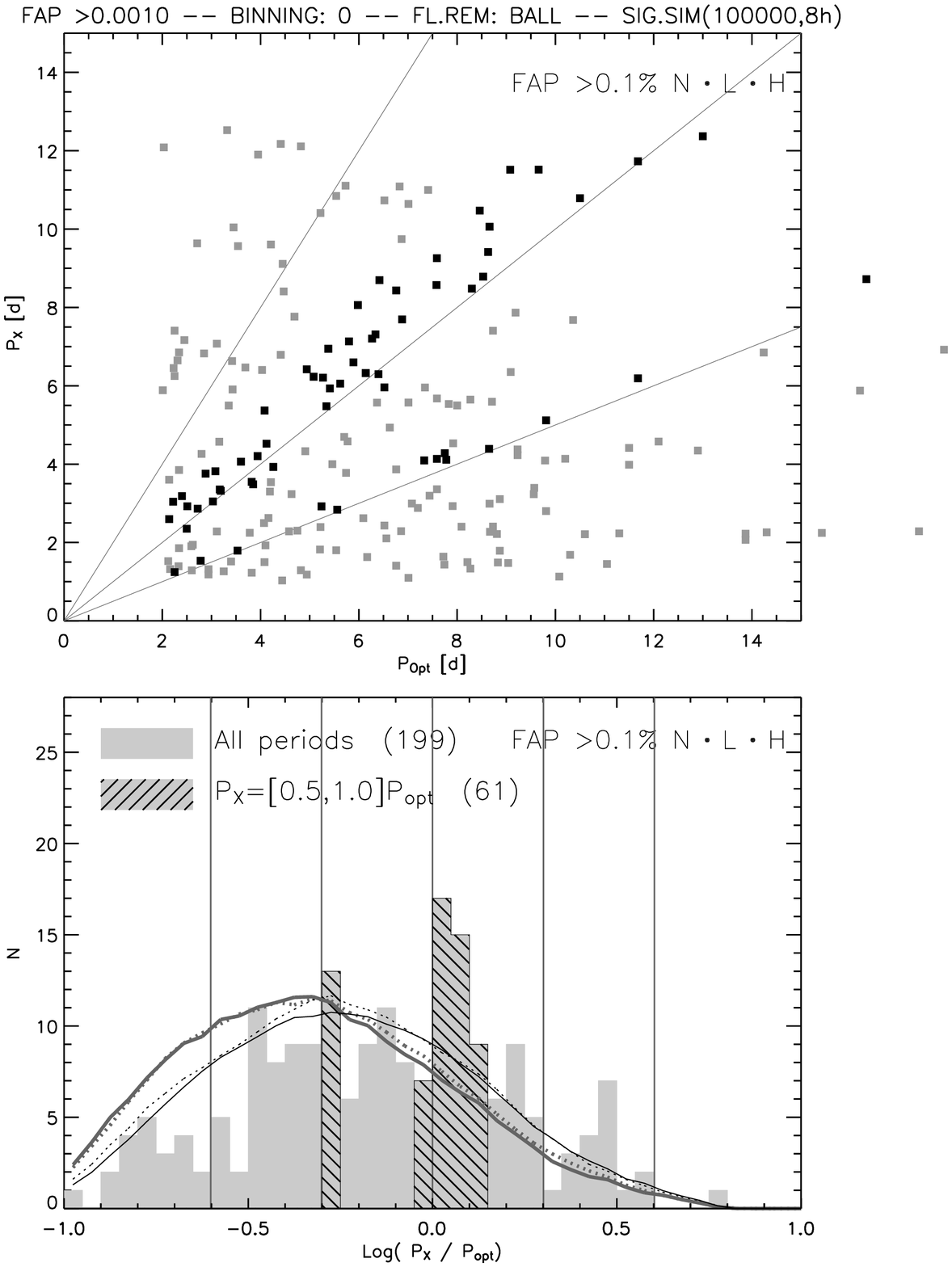}\\

 \caption{Distribution of X-ray and optical periods, as in Fig.
 \ref{fig:ratioSig}, for COUP sources without significant periodic signals
(FAP$>0.1$\%). Note that the peaks present in the previous figure have
here largely disappeared.  \label{fig:ratioOth}}

\end{figure}

\begin{figure}[!h]
\includegraphics[width=17cm]{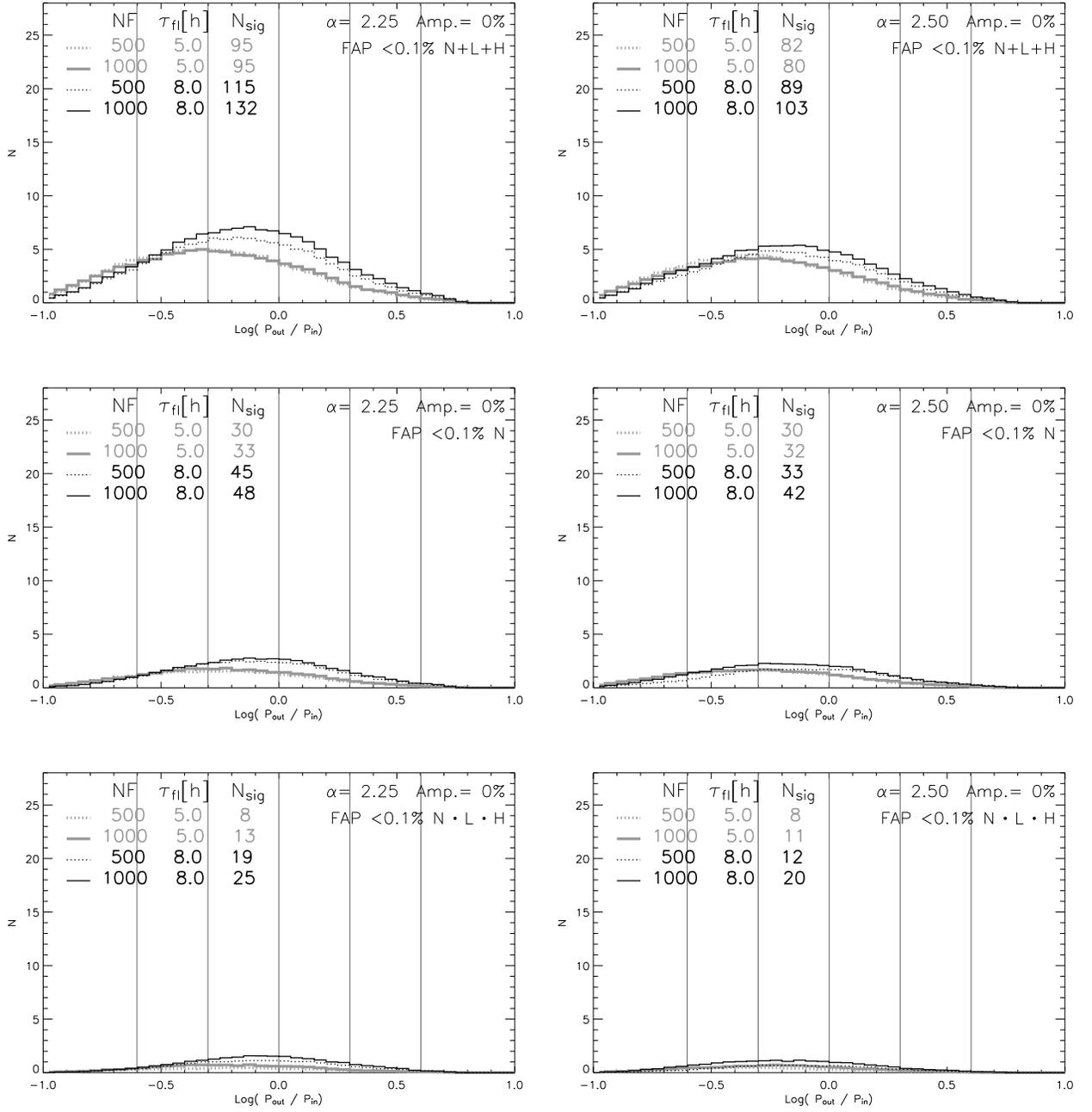}

 \caption{Histograms of $\log P_{out}/P_{in}$, resulting from the
application of our period finding method to simulated ``flaring'' light
curves (see text) with no intrinsic modulation (Amp=0\%). Rows refer to
three selections of ``significant'' periods, the same as in Figure
\ref{fig:ratioSig}. From top to bottom: ``$N+L+H$'', ``$N$'' and
``$N\cdot L\cdot H$''. The left and right columns refer to simulations
with $\alpha=2.25$ and $\alpha=2.50$.  Within each panel four
histograms are shown, for NF=500 and 1000 and for $\tau_{fl}=$5 and 8
hours, as indicated in the legend on the upper left of each panel. All
histogram are normalized so to yield the expected distributions for a
sample of 233 stars with $P_{in}$ equal to $P_{opt}$ in our
``searched'' sample. In the legend we also report the integral of these
distributions, $\rm N_{sig}$, indicating for each set of input
parameters the total number of stars expected to pass the significance
selections.  \label{fig:simA00}}

\end{figure}

\begin{figure}[!h]
\includegraphics[width=17cm]{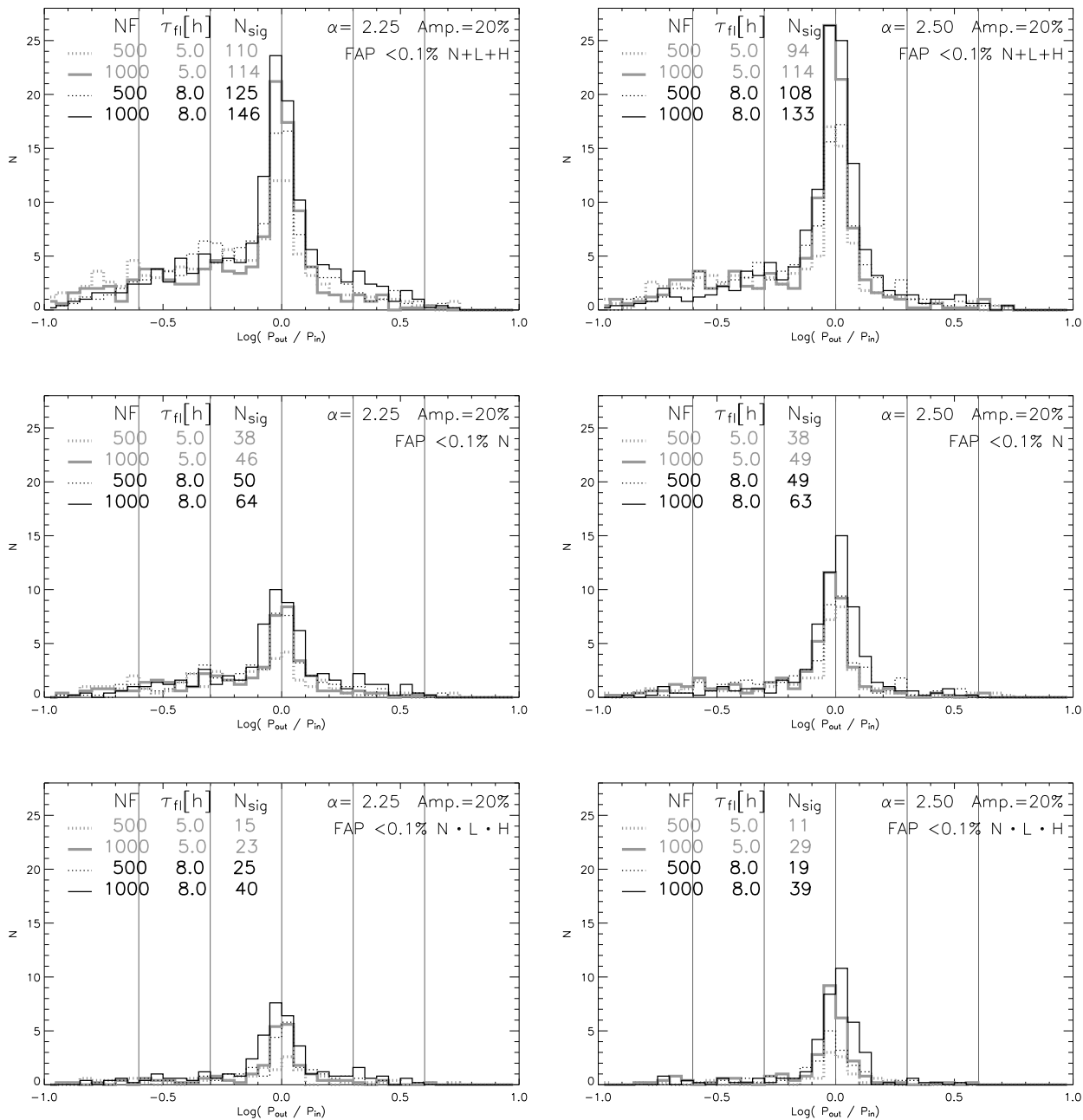}

 \caption{Same as Figure \ref{fig:simA00}, but for simulated
lightcurves with intrinsic 20\% relative modulation.
\label{fig:simA20}}

\end{figure}

Figure \ref{fig:simA20} shows in the same format results for simulated
lightcurves with a small, 20\%, intrinsic modulation. Success in
retrieving the input periods would be indicated by a narrow peak at
$\log P_{out}/P_{in}=0$ containing the largest possible fraction of the
233 input sources.  We find that most periods can be retrieved using
our method, most cleanly in the ``$N\cdot L\cdot H$'' selection that
includes heavy trimming of flares.  Tails of spurious periods are
found. Note that it is easier to retrieve the input periods for the
larger value of $\alpha$ (i.e. when large flares are less frequent) and
for the larger number of flares, $NF$ (i.e. when the total source
counts are higher). Larger relative modulation amplitudes (not shown)
also results in higher success rates. Several other indications can be
read from Figure  \ref{fig:simA20} as well as from similar ones for the
other amplitudes: \begin{enumerate}

\item The uncertainties in the derived period, i.e. the width of the peak
in the $P_{out}/P_{in}$ ratio histograms are of the same order of the
width in the corresponding plot based on real COUP data.

\item Aliases of the input periods are {\em not} seen in the simulations,
such as the $P_X = 0.5 P_{opt}$ secondary peak clearly seen in the real
COUP data. We only obtain
a peak at $P_{out}/P_{in}=1$ and a scattered tail in the range
$0.2 < P_X/P_{opt} < 5$.

\item  No systematic shift between $P_{out}$ and $P_{in}$ is observed.

\item  Short periods are determined more precisely (not shown). This occurs
because more modulation periods are covered by the
observation. It might also explain the smaller width of the
$P_X/P_{opt}\sim0.5$ peak in Figure \ref{fig:ratioSig}.

\item  Periods from $\sim 1$ day and $\sim 13$ days can be retrieved,
confirming the results of the purely periodic simulations (\S
\ref{sect:dataana}). 

\end{enumerate}

\clearpage
\newpage

\begin{deluxetable}{rrrrrrrrrrr}
\tabletypesize{\scriptsize}
\tablecaption{Data \label{tab:sample}}
\tablewidth{0pt}
\tablehead{
\colhead{COUP} & 
\colhead{Cts\tablenotemark{a}} & 
\colhead{Id.} & 
\colhead{$P_{opt}$ [d]} & 
\colhead{Ref.\tablenotemark{b}} & 
\colhead{Mass [$M_\odot$]} & 
\colhead{$\log$ Age [yr]} & 
\colhead{$\rm \log L_X$ [ergs/s]} & 
\colhead{$\rm \log L_X/L_{bol}$} & 
\colhead{EW(CaII)} & 
\colhead{Mod.Fl.\tablenotemark{c}} 
}
\startdata
      6&    1887&      40&      9.81&        H&    0.23&     6.3&    29.8&    -3.2&     0.0&      LH\\
     17&    1083&      63&      4.10&       cs&    0.90&     6.1&    30.1&    -3.8&     1.9&       -\\
     20&     382&      70&      1.50&        H&    0.16&     5.3&    29.4&    -3.7&     0.0&    \dag\\
     23&   55449&      75&      3.45&        h&    2.17&     6.2&    31.3&    -3.2&      --&       -\\
     27&    5948&      77&      1.50\tablenotemark{d}&     cshH&    0.53&     6.2&    30.3&    -3.1&     1.8&    \dag\\
     28&   20863&      81&      4.41&        H&    0.53&     6.0&    30.9&    -2.6&     1.6&       L\\
     29&    2287&      83&      7.72&       sh&    0.33&     6.1&    30.0&    -3.3&    -9.2&       -\\
     30&     169&      84&      2.45&        H&    0.25&     6.4&    28.8&    -4.1&     0.0&       -\\
     41&     574&      98&      2.34&        H&    0.21&     6.0&    29.3&    -3.7&     0.0&       -\\
     47&     837&     106&      1.70&        H&    0.27&     5.6&    29.5&    -3.8&     1.5&    \dag\\
\enddata
\tablecomments{Sample rows: the full table is available in the
electronic edition of \apj. Explanation of columns: (1) COUP source
number; (2) net, background subtracted, ACIS counts; (3) optical
source identification \citep{H97,herb02}; (4) rotation period; (5)
reference for the rotation period; (6) stellar mass; (7) log stellar
age; (8) log $\rm L_X$; (9) log $\rm L_X/L_{bol}$; (10) CaII line
equivalent width; (11) X-ray modulation flag.
}
\tablenotetext{a}{Net photons collected during the 850ksec COUP observation in the source extraction area, including  from 1\% to 96\% of the source point spread function ($\sim 90$\% for most sources; \citealt{get05}).}
\tablenotetext{b}{Reference for $P_{opt}$: H=\citet{herb02},
h=\citet{herb00}, s=\citet{stas99}, c=\citet{carp01}, U=Herbst
(unpublished)}
\tablenotetext{c}{Indicates filtered light curves (N,L or H) for which
the LNP analysis yielded FAP$<$0.1\%. A \dag \ indicates that the
periodogram analysis was not performed because either $\rm P_{opt}<2.0$d or $\rm Cts<100$.}
\tablenotetext{d}{other period:        2.99(c)}
\end{deluxetable}

\begin{deluxetable}{rrrrrrrrrrrrrr}
\tabletypesize{\scriptsize}
\tablecaption{Results of the period analysis -- ``$N\cdot L\cdot H$'' sample \label{tab:NLH}}
\tablewidth{0pt}
\tablehead{
\colhead{COUP} & 
\colhead{Cts\tablenotemark{a}} & 
\colhead{Id.} & 
\colhead{$P_{opt}$} & 
\colhead{Ref.\tablenotemark{b}} & 
\colhead{$\rm P_X^N$} & 
\colhead{$\rm FAP$} & 
\colhead{$\rm P_X^{L}$} & 
\colhead{$\rm FAP^{L}$} & 
\colhead{$\rm P_X^{H}$} & 
\colhead{$\rm FAP^{H}$} & 
\colhead{B} & 
\colhead{Amp.} & 
\colhead{Class} 
}
\startdata
     62&    9294&     123&      6.63&        s&      8.13&      0.06&      9.09&     0.00&     11.51&     0.03&  L&     0.72&                                       1\\
    131&    9038&     187&     14.41&        H&      7.92&      0.01&      7.70&     0.01&      1.97&     0.01&  N&     0.71&                              $\onehalf$\\
    139&    6094&     192&      9.04&       sH&      4.87&      0.00&      4.91&     0.00&      4.91&     0.00&  N&     0.50&                              $\onehalf$\\
    161&    2428&     211&      5.46&       sH&      5.71&      0.02&      5.92&     0.04&      5.79&     0.01&  H&     0.18&                                       1\\
    174&    2810&     222&      5.17&      shH&      4.72&      0.03&      4.66&     0.00&      4.68&     0.00&  H&     0.33&                                       1\\
    226&    2676&     258&     10.98&       cH&     12.10&      0.01&     11.91&     0.01&     10.66&     0.05&  N&     0.58&                                       1\\
    250&     497&     278&      6.76&        H&      6.39&      0.00&      6.82&     0.00&      6.71&     0.00&  H&     0.53&                                       1\\
    271&     189&     292&      5.11&        c&      2.68&      0.04&      2.33&     0.01&      2.33&     0.00&  H&     0.66&                                  $\neq$\\
    292&    1580&      --&      7.83&        c&     10.39&      0.00&     10.00&     0.00&      9.83&     0.00&  L&     0.61&                                       1\\
    413&    3204&     362&      2.73&       hH&      6.79&      0.06&      6.72&     0.01&      6.09&     0.02&  L&     0.47&                                  $\neq$\\
    454&   17125&     373&      9.81&        H&      3.91&      0.01&      3.92&     0.01&      3.99&     0.05&  N&     0.53&                                  $\neq$\\
    612&    2305&    435a&     10.33&        c&      2.98&      0.02&      3.02&     0.01&      3.03&     0.00&  H&     0.20&                                  $\neq$\\
    697&    5984&     470&     10.70&        h&      9.62&      0.00&      9.93&     0.00&     10.22&     0.00&  H&     0.38&                                       1\\
   1023&    4852&    9250&      2.27&        H&     12.18&      0.03&     11.92&     0.03&     11.63&     0.02&  H&     0.26&                                  $\neq$\\
   1070&    4733&     634&      5.38&        H&      3.53&      0.05&      3.61&     0.01&      3.61&     0.00&  H&     0.28&                                  $\neq$\\
   1116&   52865&     660&      6.15&        h&      6.64&      0.02&      6.64&     0.02&      6.64&     0.02&  H&     0.26&                                       1\\
   1141&     874&    9280&      7.92&        H&      4.12&      0.02&      4.18&     0.01&      4.20&     0.01&  L&     0.45&                              $\onehalf$\\
   1161&    9270&     690&      3.90&        H&      3.43&      0.02&      9.53&     0.00&      9.83&     0.00&  L&     0.46&                                  $\neq$\\
   1216&     637&     721&      2.45&        h&      6.55&      0.05&      6.76&     0.04&      6.63&     0.08&  L&     0.49&                                  $\neq$\\
   1233&    1056&     727&      6.03&        h&      8.40&      0.01&      8.20&     0.00&      8.20&     0.04&  L&     0.58&                                       1\\
   1248&   14567&     733&      3.28&       sH&      6.16&      0.03&      6.15&     0.03&      6.15&     0.06&  L&     0.36&                                  $\neq$\\
   1335&    5466&     782&     10.22&        c&      5.69&      0.03&      5.55&     0.02&      5.56&     0.00&  H&     0.33&                              $\onehalf$\\
   1355&    5878&     798&     10.36&        H&     11.96&      0.01&     12.14&     0.01&      5.19&     0.07&  L&     0.47&                                       1\\
   1382&   10217&    811a&     11.10&        h&      5.61&      0.00&      5.61&     0.00&      5.63&     0.00&  H&     0.25&                              $\onehalf$\\
   1384&   25433&     813&      2.85&     cshH&      2.79&      0.08&      2.87&     0.06&      2.87&     0.06&  H&     0.44&                                       1\\
   1398&    3037&     823&      9.00&       hH&     10.79&      0.01&     10.31&     0.00&      9.94&     0.00&  L&     0.61&                                       1\\
   1421&    6370&     836&     12.20&        h&      4.38&      0.05&      4.39&     0.02&      4.39&     0.04&  L&     0.26&                                  $\neq$\\
   1423&    3553&     838&      7.75&        H&      6.88&      0.06&      6.87&     0.01&      6.85&     0.00&  H&     0.29&                                  $\neq$\\
   1429&    5527&     839&      7.52&       hH&      8.22&      0.01&      8.09&     0.00&      8.04&     0.00&  L&     0.30&                                       1\\
   1448&    3304&     860&      6.33&     cshH&      7.82&      0.06&      7.89&     0.00&      7.97&     0.00&  H&     0.30&                                       1\\
   1463&    8214&     867&     10.66&        H&     11.20&      0.09&      3.95&     0.10&     10.87&     0.07&  H&     0.34&                                       1\\
   1500&    4438&     892&      8.56&       hH&     10.40&      0.00&     10.43&     0.00&     10.90&     0.00&  L&     0.50&                                       1\\
   1570&    4146&     962&      9.56&       cH&      4.99&      0.00&      4.97&     0.00&      4.92&     0.01&  L&     0.63&                              $\onehalf$\\
   1590&    1190&     982&      7.10&        c&      3.59&      0.02&      3.62&     0.01&      3.49&     0.01&  L&     0.50&                              $\onehalf$\\
\enddata

\tablecomments{Explanation of columns: (1) COUP source
number; (2) net, background subtracted,  ACIS counts; (3) optical
source identification \citep{H97,herb02}; (4) rotation period; (5)
reference for the rotation period; (6, 8, 10) best X-ray period from the
``N'', ``L'' and ``H'' analyses; (7, 9, 11) FAPs for the best X-ray
periods in ``N'', ``L'' and ``H'' analyses; (12) analysis that yielded
the lowest FAP; (13) relative amplitude of the modulation computed
from the averaged folded lightcurve as (max-min)/(max+min); (14) source
class according to $P_X/P_{opt}$.
}
\tablenotetext{a}{Net photons collected during the 850ksec COUP observation in the source extraction area, including  from 1\% to 96\% of the source point spread function ($\sim 90$\% for most sources; \citealt{get05}).}
\tablenotetext{b}{Reference for $P_{opt}$: H=\citet{herb02},
h=\citet{herb00}, s=\citet{stas99}, c=\citet{carp01}}
\end{deluxetable}

\clearpage
\newpage

\clearpage
\newpage


\clearpage
\newpage

\clearpage

\clearpage

\clearpage

\clearpage
\newpage

\clearpage
\newpage

\begin{figure}[!t]
\begin{center}
\includegraphics[width=16cm]{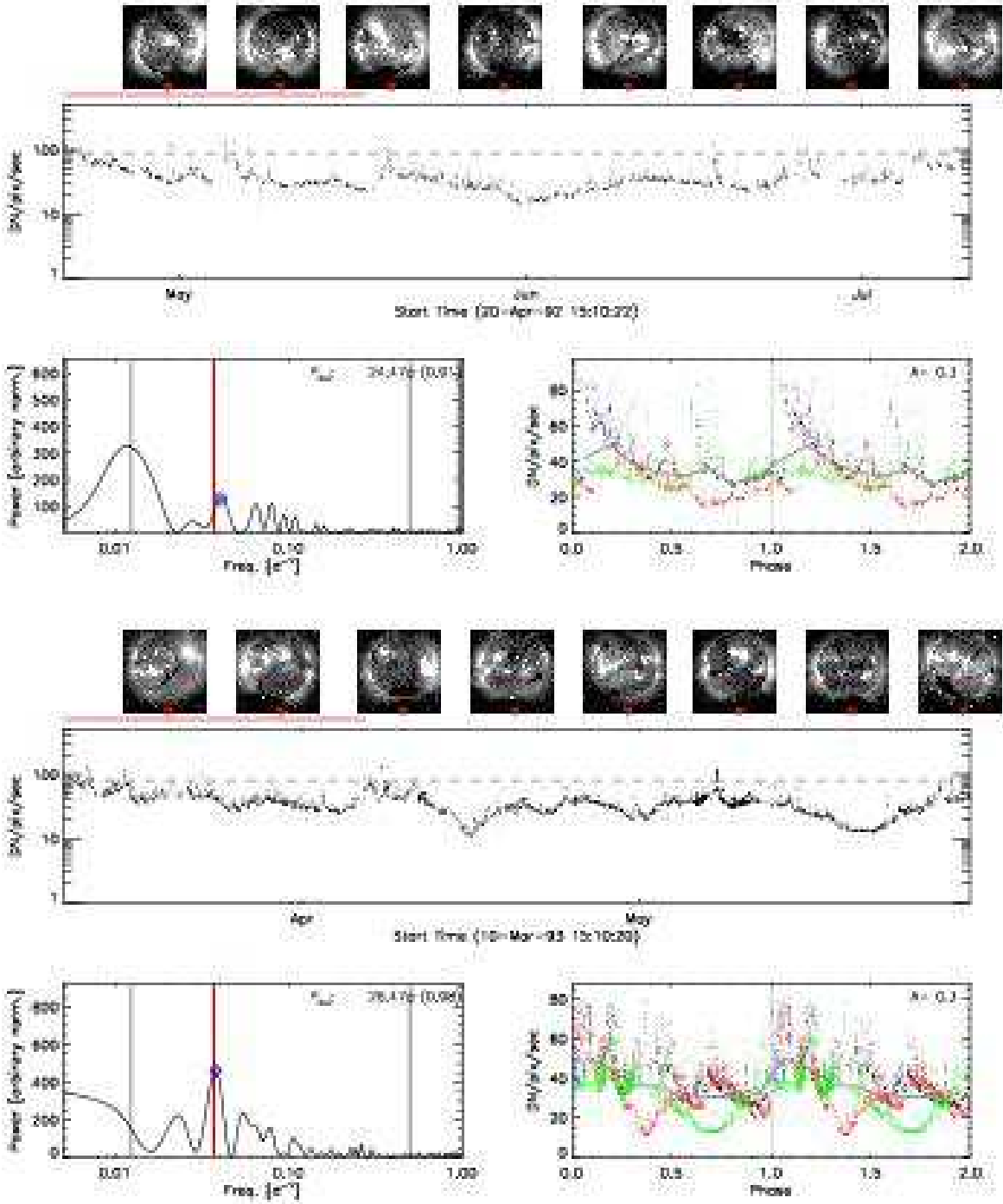} \\

 \caption{Same as figure \ref{fig:sun_lnp}, for two other time segments
 both close to solar maximum.  \label{fig:sun_lnp2}}
\end{center}
\end{figure}

\clearpage
\newpage

\clearpage

\begin{figure*}[!t]
\center
\Large This figure is available in the electronic version of ApJ. A version of the paper that includes the 34 panels of this figure can be downloaded from: 

http://www.astropa.unipa.it/$\sim$ettoref/COUP\_RotMod.pdf

\vspace*{2cm}
\caption{Sources for which all filtering methods yield FAP$<$0.1\%
(``$N\cdot L\cdot H$'' selection)
\label{fig:exLC6}}
\end{figure*}

\end{document}